%% file: fse26.tex
\documentclass[acmsmall]{acmart}
\AtBeginDocument{%
  }

\newcommand{\para}[1]{\vspace{2pt}\noindent\textbf{#1.~}}
\newcommand{\ignore}[1]{}
\newcommand{\system}{\textit{\sloppy{V2E}\@}}

\usepackage{algorithmic}
\usepackage{graphicx}
\usepackage{textcomp}
\usepackage[table,xcdraw]{xcolor}
\usepackage{array}

\usepackage{algorithm}
\usepackage{tcolorbox}
\input{samples/solidity}
\input{samples/vyper}
\usepackage{graphicx}
\usepackage{subcaption}
\usepackage{url}
\usepackage{hyperref}
\usepackage{booktabs}
\usepackage{multirow}
\graphicspath{{picture}}
\usepackage{enumitem}
\usepackage{dashrule}
\usepackage{amsmath}
\usepackage{longtable}

\renewcommand{\algorithmicrequire}{ \textbf{Input:}}
\renewcommand{\algorithmicensure}{ \textbf{Output:}}

\setcopyright{rightsretained}
\acmDOI{10.1145/3808108}
\acmYear{2026}
\acmJournal{PACMSE}
\acmVolume{3}
\acmNumber{FSE}
\acmArticle{FSE101}
\acmMonth{7}
\acmSubmissionID{fse26mainb-p150-p}
\received{2025-09-11}
\received[accepted]{2026-03-24}

\begin{document}
% \title{AGen: Feedback and Profit Driven Auto Exploit Generation for Smart Contract Vulnerability Verification}
%\title{AGen: Exploit and Profit Driven Automatic Smart Contract Vulnerability Severity Verification}

\title{V2E: Validating Smart Contract Vulnerabilities through Profit-driven Exploit Generation and Execution}



\author{Jingwen Zhang}
\orcid{0009-0002-8611-3120}
\affiliation{%
  \institution{Sun Yat-sen University and Peng Cheng Laboratory}
  % \city{Shenzhen}
  \country{China}}
\email{zhangjw273@mail2.sysu.edu.cn}

\author{Yuhong Nan}
\orcid{0000-0001-9597-9888}
\authornote{Yuhong Nan is the corresponding author.}
\email{nanyh@mail.sysu.edu.cn}
\affiliation{%
  \institution{Sun Yat-sen University}
  % \city{Zhuhai}
  \country{China}
}

\author{Kaiwen Ning}
\orcid{0009-0009-6009-8285}
\affiliation{%
  \institution{Sun Yat-sen University and Peng Cheng Laboratory}
  % \city{Shenzhen}
  \country{China}}
\email{ningkw@mail2.sysu.edu.cn}

\author{Mingxi Ye}
\orcid{0009-0004-6708-4074}
\affiliation{%
  \institution{Sun Yat-sen University}
  % \city{Zhuhai}
  \country{China}}
\email{yemx6@mail2.sysu.edu.cn}

\author{Wei Li}
\orcid{0009-0002-1333-4408}
\affiliation{%
  \institution{Sun Yat-sen University}
  % \city{Zhuhai}
  \country{China}}
\email{liwei378@mail2.sysu.edu.cn}

\author{Yuming Xiao}
\orcid{0009-0004-4914-5136}
\affiliation{%
  \institution{Sun Yat-sen University}
  % \city{Zhuhai}
  \country{China}}
\email{xiaoym23@mail2.sysu.edu.cn}

\author{Yuming Feng}
\orcid{0000-0001-8922-0496}
\affiliation{%
  \institution{Peng Cheng Laboratory}
  % \city{Shenzhen}
  \country{China}}
\email{fengym@pcl.ac.cn}

\author{Weizhe Zhang}
\orcid{0000-0003-4783-876X}
\affiliation{%
  \institution{Harbin Institute of Technology}
  \country{China}}
\email{wzzhang@hit.edu.cn}

\author{Zibin Zheng}
\orcid{0000-0002-7878-4330}
\affiliation{%
  \institution{Sun Yat-sen University}
  % \city{Zhuhai}
  \country{China}}
\email{zhzibin@mail.sysu.edu.cn}
\renewcommand{\shortauthors}{Jingwen Zhang et al.}

\begin{abstract}
  Smart contracts are a critical component of blockchain systems. Due to the large amount of digital assets carried by smart contracts, their security is of critical importance. Although numerous tools have been developed for detecting smart contract vulnerability, their effectiveness remains limited, particularly due to the high false positives included in the reported results. Therefore, developers and auditors are often overwhelmed with manually verifying the reported issues. A fundamental reason behind this is that while a reported vulnerability satisfies specific vulnerable patterns, it may not actually be exploitable, either because the vulnerable code cannot be triggered or it does not result in any financial loss.
  
  % significant number of false positives in their results. Specifically, most of these tools either cannot confirm the reachability of the vulnerability or fail to assess the vulnerability's severity, such as whether it is profitable. As a result, developers are often overwhelmed with manually verifying the reported issues, resulting in the continued occurrence of contract attacks and significant financial losses.
  
  % most of them cannot adapt to contract evolution, fail to handle dynamically determined exploit severity, and rely on conservative strategies to report vulnerabilities. As a result, developers and auditors are unable to efficiently evaluate the real-world impact of detected issues. Moreover, this limitation results in the continued occurrence of contract attacks and significant financial losses. Therefore, to support rapid vulnerability response and remediation, vulnerability validation is essential. 

    % Smart contract vulnerability detection important, due to its close-relevance to digital assets/profits/exploits
    % While a number of detection tools exist, the reported results are less satisfied - (1) vulnerable does not means exploitable, (2) false alarms exists, due to approach-wise limitations.
    % Confirming those severity mostly rely on manual analysis/human efforts.

    % Our work, a new approach to make things a step further -- validating if a reported vulnerability is truly exploitable. The core idea - (1) profit-driven, (2) validate/find the real-path that can trigger the vulnerability.
  
  In this paper, we propose \system{}, a new framework for validating whether a reported vulnerability is truly exploitable. The core idea of \system{} is to automatically generate executable Proof-of-Concept Exploit~(PoC for short), and then assess if the vulnerability could be triggered and incur any real damage (i.e., causing financial loss) by the PoC. While LLMs have shown proficiency in PoC generation, achieving our task is by no means trivial. In detail, it is difficult for LLM to: (1) generate and update PoC to trigger a specific vulnerability, (2) evaluate the PoC's effectiveness to validate exploitable vulnerability. To this end, \system{} automates the whole process through a novel combination of PoC generation, validation, and refinement: (1) Firstly, \system{} generates targeted PoCs by analyzing potential vulnerability paths. (2) Then, \system{} verifies the validity of PoCs through triggerability and profitability analysis. (3) In addition, \system{} iteratively refines the generated PoC based on PoC execution feedback, therefore, increasing the chance to confirm the vulnerability. Evaluation on 264 manually labeled contracts shows that \system{} outperforms the baseline approach. Particularly, \system{} successfully identifies 102 out of 124 exploitable vulnerabilities, achieving a precision of 91.9\% and a recall of 82.3\%. In addition, it successfully eliminates 71 out of 140 false alarms (50.7\%). Besides, \system{} effectively enhances the performance of SOTA tools. It reduces the false positive rates of Slither by 76.9\%, Mythril by 56.9\% and Confuzzius by 65\%.
  % we propose an LLM and program analysis-based framework, called \system{},  to automatically construct executable proof-of-concept (PoC) to validate the severity of smart contract vulnerabilities.
  % Evaluation of 264 vulnerable contracts shows that \system{} achieves a precision and recall rate of 91.9\% and 82.3\% respectively, identifies 102 out of 124 vulnerabilities, and successfully eliminates 71 out of 140 false alarms (50.7\%) \yuhong{compared to XX}.
\end{abstract}

\begin{CCSXML}
<ccs2012>
   <concept>
       <concept_id>10011007.10011074.10011099.10011102.10011103</concept_id>
       <concept_desc>Software and its engineering~Software testing and debugging</concept_desc>
       <concept_significance>500</concept_significance>
       </concept>
 </ccs2012>
\end{CCSXML}

\ccsdesc[500]{Software and its engineering~Software testing and debugging}

\keywords{Smart Contract, Vulnerability Validation, Proof of Concept}


\maketitle

%% main content
\input{content}

\begin{acks}
This work was supported in part by the National Natural Science Foundation of China (No. 62572497, No. 624B2139), NSFC-RGC Collaborative Research (No. 62461160332), Guangdong Zhujiang Talent Program (No. 2023QN10X561), and the Major Key Project of Pengcheng Laboratory under Grant PCL2025A07.
\end{acks}

\bibliographystyle{samples/ACM-Reference-Format}
\bibliography{reference}

% \appendix

\end{document}

%% file: samples/solidity.tex
\usepackage{listings, xcolor}

\definecolor{verylightgray}{rgb}{.97,.97,.97}

\lstdefinelanguage{Solidity}{
	keywords=[1]{anonymous, assembly, assert, break, call, callcode, case, catch, class, constant, continue, constructor, contract, debugger, default, delegatecall, delete, do, else, emit, event, experimental, export, external, false, finally, for, function, gas, if, implements, import, in, indexed, instanceof, interface, internal, is, length, library, log0, log1, log2, log3, log4, memory, modifier, new, payable, pragma, private, protected, public, pure, push, require, return, returns, revert, selfdestruct, send, solidity, storage, struct, suicide, super, switch, then, this, throw, transfer, true, try, typeof, using, value, view, while, with, addmod, ecrecover, keccak256, mulmod, ripemd160, sha256, sha3}, % generic keywords including crypto operations
	keywordstyle=[1]\color{blue}\bfseries,
	keywords=[2]{address, bool, byte, bytes, bytes1, bytes2, bytes3, bytes4, bytes5, bytes6, bytes7, bytes8, bytes9, bytes10, bytes11, bytes12, bytes13, bytes14, bytes15, bytes16, bytes17, bytes18, bytes19, bytes20, bytes21, bytes22, bytes23, bytes24, bytes25, bytes26, bytes27, bytes28, bytes29, bytes30, bytes31, bytes32, enum, int, int8, int16, int24, int32, int40, int48, int56, int64, int72, int80, int88, int96, int104, int112, int120, int128, int136, int144, int152, int160, int168, int176, int184, int192, int200, int208, int216, int224, int232, int240, int248, int256, mapping, string, uint, uint8, uint16, uint24, uint32, uint40, uint48, uint56, uint64, uint72, uint80, uint88, uint96, uint104, uint112, uint120, uint128, uint136, uint144, uint152, uint160, uint168, uint176, uint184, uint192, uint200, uint208, uint216, uint224, uint232, uint240, uint248, uint256, var, void, ether, finney, szabo, wei, days, hours, minutes, seconds, weeks, years},	% types; money and time units
	keywordstyle=[2]\color{teal}\bfseries,
	keywords=[3]{block, blockhash, coinbase, difficulty, gaslimit, number, timestamp, msg, data, gas, sender, sig, value, now, tx, gasprice, origin},	% environment variables
	keywordstyle=[3]\color{violet}\bfseries,
	identifierstyle=\color{black},
	sensitive=true,
	comment=[l]{//},
	morecomment=[s]{/*}{*/},
	commentstyle=\color{gray}\ttfamily,
	stringstyle=\color{red}\ttfamily,
	morestring=[b]',
	morestring=[b]"
}

%% file: samples/vyper.tex
\usepackage{listings, xcolor}

\definecolor{verylightgray}{rgb}{.97,.97,.97}

\lstdefinelanguage{Vyper}{
    keywords=[1]{@external, @private, @view, @pure, @payable, @nonreentrant, assert, break, continue, contract, create, selfdestruct, elif, else, emit, event, for, if, in, interface, internal, log, loop, struct, return, returns, raise, static, supports, while, as, unit, use, from, import, pass, True, False, None, def}, % 控制流和修饰符
    keywordstyle=[1]\color{blue}\bfseries,
    keywords=[2]{address, bool, bytes32, decimal, int128, int256, uint8, uint256, string, Bytes, HashMap, timestamp, timedelta, wei_value, currency_value,} % 数据类型和单位
    keywordstyle=[2]\color{teal}\bfseries,
    keywords=[3]{block, chain, msg, tx, chain.id, chain.coinbase, block.timestamp, block.number, block.difficulty, block.gas_limit, msg.sender, msg.value, msg.data, msg.gas, tx.origin, tx.gasprice, self}, % 环境变量
    keywordstyle=[3]\color{violet}\bfseries,
    keywords=[4]{send, raw_call, create_forwarder_to, create_copy_of, empty, len, slice, concat, keccak256, ecadd, ecmul, ecpairing, extract32, as_wei_value, as_unit_number, min, max, shift, pow}, % 内置函数
    keywordstyle=[4]\color{orange}\bfseries,
    identifierstyle=\color{black},
    sensitive=true,
    comment=[l]{\#},
    morecomment=[s]{'''}{'''},
    commentstyle=\color{gray}\ttfamily,
    stringstyle=\color{red}\ttfamily,
    morestring=[b]'',
    morestring=[b]"
}

%% file: content.tex
\section{Introduction}
Smart contracts are self-executing code and form a critical component of blockchain systems. Due to their inherent financial properties, smart contracts have become prime targets for attackers. For example, the notorious DAO attack~\cite{dao} resulted in losses exceeding 50 million USD. To secure smart contracts, numerous studies have proposed effective vulnerability detection techniques, such as static analysis~\cite{astro, gigahorse}, fuzzing~\cite{verite, sFuzz}, and machine learning based methods~\cite{llm4fuzz, iaudit}. 

% However, prior research~\cite{chaliasos2024smart} has shown that developers and auditors still tend to prefer manual vulnerability audit. Firstly, existing tools produce excessive false alarms~\cite{ghaleb2020effective}, as they cannot adapt to the continuous evolution of smart contracts and rely heavily on predefined rules to detect vulnerabilities. Besides, current tools fail to explore complex execution for exploit severity analysis~\cite{fuzz_review}, due to the low semantic nature of bytecode. Finally, detection tools tend to adopt conservative strategies, labeling all potential vulnerabilities as harmful~\cite{hu2024smart}. However, when a smart contract vulnerability path is only technically exploitable, the vulnerability can be considered nonexistent, as it does not lead to any tangible impacts. For example, if a Reentrancy can be successfully executed but does not result in any profit, attackers are unlikely to exploit it, since their primary motivation is financial gain~\cite{nyx}. As a result, developers and auditors need to manually construct and evaluate test cases, which hinders their ability to prioritize and address vulnerabilities. 

% However, existing studies[xx] —show that current smart contract vulnerability detectors suffer from a high rate of false positives, limiting their practical usability. In particular, many detected vulnerabilities may not be exploitable in real scenarios. Firstly, there might not be a feasbile path to trigger the vulnerability, due to the lack of approipate execution context.

However, prior studies~\cite{chaliasos2024smart, hu2024smart} have shown that existing smart contract vulnerability detection tools often suffer from high false positive rates, limiting their practical usability. More specifically, many of the reported vulnerabilities are not truly exploitable in real-world scenarios. One key reason is the absence of an appropriate execution context to trigger the vulnerability. Since these tools typically rely on static analysis or predefined patterns~\cite{ghaleb2020effective}, they struggle to capture contextual information necessary for exploiting the vulnerability. Moreover, even when the vulnerability can be triggered, it could still be unexploited if it does not lead to any tangible damage. In other words, if a vulnerability can be successfully triggered but does not result in any profit, attackers are unlikely to exploit it, since their primary motivation is financial gain~\cite{nyx}. 
%Most existing tools are unable to assess the real impact of an exploit because they cannot handle low-semantic execution bytecode~\cite{fuzz_review}.  As a result, developers and auditors still need to manually construct and evaluate test cases, which hinders their ability to prioritize and fix vulnerabilities.

% Determining whether a vulnerability path is reachable requires consideration of the execution context, while these tools fail to capture it due to their reliance on predefined rules~\cite{ghaleb2020effective}. 

% This affects the ability to prioritize and respond effectively to truly harmful vulnerabilities.
% the speed of auditing cannot keep up with the rapid growth of contracts, leaving attacks uncontrolled.

% existing tools suffer from high false positive rates~\cite{hu2024smart}. They often rely heavily on predefined rules to detect vulnerabilities, overlooking the actual semantics and intent of the code. As a result, they generate excessive false alarms, which fail to reduce the manual workload. On the other hand, most detection methods ignore the actual impact of the vulnerabilities~\cite{fuzz_review}. Developers, auditors, and attackers are primarily concerned with vulnerabilities that can lead to tangible losses. Nevertheless, current tools lack the capability for fine-grained analysis of severity, such as whether a vulnerability can be used to gain financial profit, forcing developers to manually construct test cases to validate the real-world impact of detected issues.
% 

Therefore, helping developers and auditors automatically verify the reported vulnerabilities is critically important, as it significantly accelerates vulnerability response and remediation. To achieve this goal, one promising approach is to automatically generate the Proof-of-Concept Exploit (called PoC), a code snippet that is used to prove the validity of the reported vulnerability. In detail, PoCs provide both an executable script to trigger the vulnerability and concrete evidence of its impact. Besides, recent advancements~\cite{faultline, pocgen} have shown LLMs perform well in PoC generation.

% existing tools suffer from high false positive rates~\cite{hu2024smart}. They often rely heavily on predefined rules to detect vulnerabilities, overlooking the actual semantics and intent of the code. As a result, they generate excessive false alarms, which fail to reduce the manual workload. On the other hand, most detection methods ignore the actual impact of the vulnerabilities~\cite{fuzz_review}. Developers, auditors, and attackers are primarily concerned with vulnerabilities that can lead to tangible losses. Nevertheless, current tools lack the capability for fine-grained analysis of exploit severity, such as whether a vulnerability can be used to gain financial profit, forcing developers to manually construct test cases, e.g., Proof-of-Concept~(PoC), to validate the real-world impact of detected issues.

\para{Challenges}
% To validate the vulnerability through PoC, assessing the PoC’s effectiveness is crucial. However, although LLMs can directly generate PoCs, evaluating PoCs' validity remains difficult. Specifically, there are two main challenges:
However, most smart contract vulnerabilities are state-dependent logic bugs, and the exploitation primarily results in inconsistency from intended execution rather than program crashes. This makes it difficult for LLMs to generate valid PoCs for smart contracts. Specifically, there are two main challenges:

% While generating PoCs with LLM is a feasible method for verifying smart contract vulnerability severity, an effective PoC should not only demonstrate the existence of the vulnerability but also enable the assessment of its impact. 

% $\bullet$ \textbf{Generating and refining vulnerability-specific PoC.} LLMs struggle to generate tailored PoCs without sufficient information. On one hand, different vulnerabilities exhibit distinct characteristics, such as nested calls in Reentrancy, reliance on block attributes in Randomness from Chain Attributes, and inconsistent execution results caused by Transaction Order Dependence. On the other hand, most static analysis~\cite{slither, defiTainter} and machine learning methods~\cite{gptscan,xfuzz} typically can only pinpoint the locations of vulnerabilities (e.g., code segments) but fail to provide the complete exploiting paths.

$\bullet$ \textbf{Generating and refining vulnerability-specific PoC.} The reasoning capabilities of LLMs are insufficient for directly generating effective PoCs. On the one hand, most static analysis~\cite{slither, defiTainter} and machine learning methods~\cite{gptscan,xfuzz} typically can only pinpoint the locations of vulnerabilities (e.g., code segments). And LLMs are struggling to infer the correct exploit path from a large number of publicly available functions within a contract. On the other hand, LLMs cannot dynamically explore the contract’s state space, involving intricate transaction sequences and parameter choices. Consequently, PoCs generated by LLMs may fail to trigger the vulnerability or even fail to execute.

$\bullet$ \textbf{Assessing effectiveness of PoC.} In the meantime, LLM themselves cannot determine whether a generated PoC is effective, since smart contract vulnerabilities can only be confirmed through actual execution. Moreover, using static analysis and fuzzing methods to evaluate PoC effectiveness is still challenging, as existing techniques~\cite{smartdagger, unity,fuzz_review} either cannot handle complex processes in PoC, such as exploit preparation, exploitation, and post-processing, or assess vulnerability severity under the obfuscation introduced by PoC characteristics, like cheatcodes. In addition, recent work (i.e., A1~\cite{A1}) adopts inline commands (e.g., \texttt{forge test}) to obtain PoC outcomes. Unfortunately, A1 offers limited extensibility for PoC-centric vulnerability assessment, as it cannot capture fine-grained execution information such as opcode invocation sequences. To this end, validating and assessing smart contract PoCs remains quite difficult.

\vspace{-0.1cm}
\begin{figure}[hbp]
    \centering
    \includegraphics[width=0.8\textwidth]{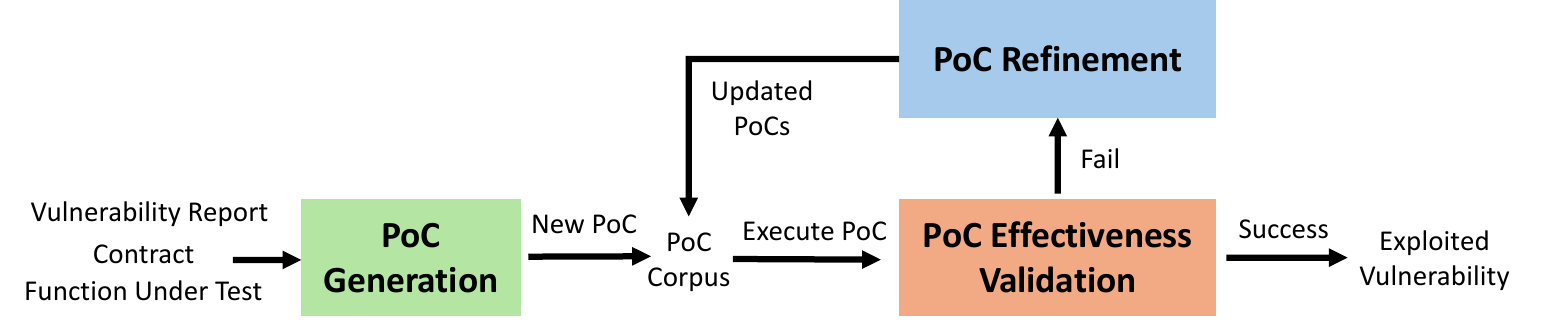}
    \caption{An example of vulnerability validating process in \system{}.}
    \label{fig:simplify_workflow}
\end{figure}
\vspace{-0.2cm}

\para{Our work}
In this paper, we introduce \system{}, a novel framework for validating smart contract vulnerabilities. As shown in Fig.~\ref{fig:simplify_workflow}, \system{} takes vulnerability report, contract source code, and function under test as input and attempts to generate PoCs to assess vulnerability. To address the challenges above, \system{} is designed as follows: (1)~analyzing vulnerability path to synthesize tailored PoC; (2)~verifying effectiveness of PoC through triggerability and profitability; (3)~using execution feedback to guide PoC update.
% tracing bytecode execution flow to explore PoC exploitability; (3) combining primitive operation to enhance PoC profitability.

Specifically, to synthesize a tailored PoC, \system{} employs fine-grained static analysis to search for vulnerability entry points, such as publicly accessible functions, and constructs potential vulnerability paths. Then, by combining vulnerability characteristics and contract deployment requirements, \system{} generates vulnerability-specific PoCs and stores them in PoC Corpus~(Section~\ref{subsec:poc_synthesis}). 
To verify the effectiveness of PoC, \system{} builds a PoC effectiveness validation framework. The framework extracts a PoC from the corpus each time and executes it. Then, \system{} analyzes whether the PoC can trigger vulnerabilities and assesses the severity of the vulnerabilities. If the PoC can both trigger the vulnerability and cause profit for attackers, \system{} considers the vulnerability is exploitable~(Section~\ref{subsec:poc_analysis}). Otherwise, \system{} refines the PoC.
To improve the opportunity of vulnerability validation, \system{} combines execution feedback for vulnerability exploration enhancement. In the case of PoC execution failure, \system{} revises the PoC using the semantics of the execution bytecode. Conversely, if the PoC runs successfully but does not trigger the vulnerability or yield profits, \system{} leverages primitive operations to refine the PoC. Then, \system{} puts the updated PoC back to the corpus for further analysis~(Section~\ref{subsec:poc_fine_tuning}).

To evaluate the effectiveness of \system{}, we collect 264 vulnerable contracts from the SmartBugs dataset, including 64 manually constructed contracts and 200 real-world on-chain contracts. We manually annotate the exploitability of each vulnerability based on whether it could be triggered for profit. Experiment results show that \system{} outperforms the LLM-based baseline method. In particular, \system{} accurately identifies 102 in 124 exploitable vulnerabilities, achieving a precision of 91.9\%, a recall of 82.3\%, and correctly filters out 71 in 140~(50.7\%) false alarms. Moreover, \system{} effectively reduces the false positive rate of Slither by 76.9\%, Mythril by 56.9\% and Confuzzius by 65\%.

In summary, this paper makes the following contributions:

\begin{itemize}

    % \yuhong{\item We highlight the need of vulnerability validation in smart contract. We propose the idea to validate vulnerability based on reachability and profitability.}

    % \yuhong{\item We propose a set of novel mechanisms to generate valid PoC, observe and interpret the results to confirm a reported vulnerability.}
    
    % \item We propose \system{}, a novel automated framework for validating smart contract vulnerabilities. \system{} innovatively combines LLMs with program analysis to enable effective PoC generation, validation, and exploration.
    % \item We design an automated PoC effectiveness verification method based on the characteristics of smart contracts. This method can effectively analyze the exploitability and profitability of PoC, and thereby validate the existence of vulnerability. 
    \item We highlight the need for vulnerability validation in smart contracts. We propose the idea to validate vulnerability based on triggerability and profitability.
    \item We propose \system{}, a novel automated smart contract vulnerability validating framework, to generate and refine vulnerability-specific PoC, observe and interpret its execution results to confirm a reported vulnerability.
    \item We conduct extensive experiments to demonstrate the effectiveness of \system{}. The results show that \system{} can effectively evaluate the vulnerabilities and enhance the performance of existing detection tools.
    % \item We release the artifact and the datasets of \system{}.
\end{itemize}

The rest of the paper is organized as follows: Section~\ref{sec:background} introduces the knowledge and motivation of our research. Section~\ref{sec:designs} presents the design choices and the workflow of \system{}. Section~\ref{sec:approach_details} outlines technology details of \system{}. Section~\ref{sec:evaluation} discusses the design and the results of experiments. Section~\ref{sec:discussion} provides a discussion and the limitations of \system{}. Section~\ref{sec:related_work} reviews related research work. Section~\ref{sec:conclusion} concludes the paper, and Section~\ref{sec:data_availability} provides details on the data availability.

% First, \system{} employs bytecode semantics to construct an execution semantic flow graph, and updates the PoC with execution flow guidance to explore exploitability. Then, \system{} refines the PoC using predefined primitive operations to generate more profits.

% \system{} employs fine-grained static analysis to search for vulnerability entry points, such as publicly accessible functions, and constructs potential vulnerability paths. The insight here is that vulnerabilities vary, but they usually require control over the contract's state to be exploited. Therefore, analyzing functions affecting states in vulnerable functions helps build exploit paths. 

\section{Background and Motivation}
\label{sec:background}

\subsection{Background}
\label{subsec:background}
\para{Smart contracts and vulnerability detection}
A smart contract~\cite{DApps} is a piece of code that runs automatically on the blockchain, like Ethereum~\cite{ethereum}. To execute a smart contract, the contract must first be compiled into bytecode and deployed onto the blockchain. Due to the large amount of funds stored in smart contracts and their immutable nature once deployed, vulnerabilities in smart contracts are a major concern for both developers and attackers. Since most vulnerabilities in smart contracts are logic-based, different vulnerabilities exhibit distinct features, such as nested calls in Reentrancy, reliance on block attributes in Randomness from Chain Attributes, and inconsistent execution results caused by Transaction Order Dependence. Although many studies~\cite{achecker, sFuzz, sailfish} have proposed vulnerability detection methods, the vast majority either cannot confirm whether a vulnerability is actually triggerable or fail to assess its real-world impact, resulting in false alarms.
% As a result, these approaches suffer from high false positives.

\para{Vulnerability validation}
Compared to vulnerability detection, vulnerability validation focuses on assessing whether a reported vulnerability can be exploited, including being triggered under real-world conditions and causing real damage. In smart contracts, this typically requires analyzing whether the vulnerability can be triggered in an on-chain environment and whether the attacker can gain profit. Besides, vulnerability validation is crucial, as it helps developers prioritize and respond to harmful vulnerabilities more efficiently under limited resources, such as time and manpower.
% While detection remains at the level of theoretical code analysis, severity validation emphasizes whether an attacker can interact with the contract’s current state to cause concrete harm, such as unauthorized asset transfers or privilege escalation.

\vspace{-0.3cm}
\begin{figure}[hbp]
    \centering
    \includegraphics[width=0.9\textwidth]{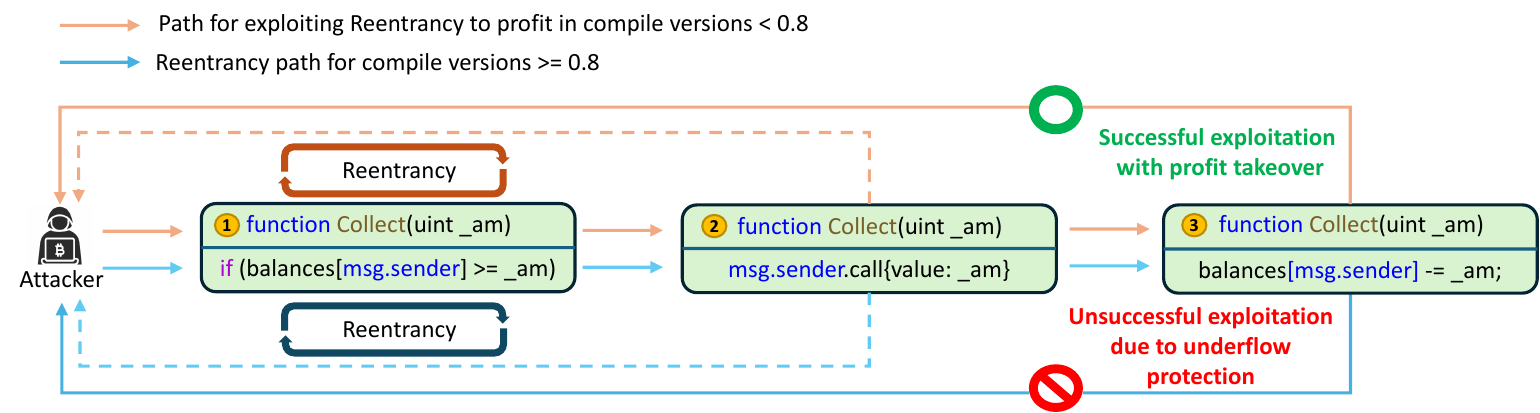}
    \caption{Motivating example. Even for the same vulnerability, different compile versions lead to different exploitability.}
    \label{fig:motivating_example}
\end{figure}
\vspace{-0.5cm}

\subsection{Motivating Example}
\label{subsec:motivating}
The function named \texttt{Collect} shown in Fig.~\ref{fig:motivating_example} contains a Reentrancy vulnerability, where an attacker can hijack the control flow at Step 2 and repeatedly withdraw funds from the function. However, what makes this example interesting is that the exploitability of the vulnerability varies depending on the compiler version. Specifically, compile versions above 0.8 automatically insert underflow checks for all arithmetic operations during compilation~\cite{soliditychanges}. As a result, in compile versions greater than 0.8, this function can no longer be exploited for profit via Reentrancy, and the Reentrancy can be considered mitigated. However, for compile versions below 0.8, the compiler does not insert additional safety checks, allowing Step 3 to execute successfully despite an underflow. The attacker can gain profit and pose a threat to the contract.

\para{Motivation of \system{}}
Despite the numerous smart contract vulnerability detection tools available, accurately identifying the exploitable vulnerability shown in Fig.~\ref{fig:motivating_example} is rather difficult. More specifically, we outline the following motivations:

$\bullet$ \textbf{Confirming the triggerability.} Most vulnerability detection tools ignore the execution context and
often rely on code patterns to ensure completeness of their results. However, this approach is inherently fragile, as it may cause benign code to be misidentified. For example, even well-maintained tools, such as Slither~\cite{slither}, Mythril~\cite{mythril}, and LLM-based methods, incorrectly identify the code in Fig.~\ref{fig:motivating_example} as vulnerable across all compile versions, failing to account for execution differences introduced by varying compiler versions. 

$\bullet$ \textbf{Evaluating exploit severity.} The severity of a detected vulnerability can only be confirmed by execution, as smart contracts are state-dependent. However, fuzzing-based approaches may misjudge the actual impact of vulnerabilities due to the low semantic nature of bytecode. For instance, at the bytecode level, \texttt{balances[msg.sender]} in Step 3 of Fig.~\ref{fig:motivating_example} is an \texttt{id} determined at runtime. Existing methods, such as Confuzzius~\cite{confuzzius}, detect the state change but are unable to interpret further what changes in \texttt{id}'s value signify. As a result, existing methods are unable to perform a fine-grained assessment of exploit severity.

\para{Key observation for exploitable smart contract vulnerability }
Through the analysis of past attacks and the results produced by vulnerability detection tools, we summarize the critical criteria that constitute exploitable smart contract vulnerability as follows:

(1) \textit{Valid trigger.} Confirming that a vulnerability can be triggered in an on-chain environment is crucial, as the same code fragment may behave differently under varying execution contexts. For example, in Step 3 at Fig.~\ref{fig:motivating_example}, when \texttt{balances[msg.sender]} is less than \texttt{\_am}, it will throw an error in compile versions above 0.8 due to built-in underflow checks, but will execute successfully in versions below 0.8. Such fine-grained behavioral differences can only be demonstrated by appropriate execution contexts.

(2) \textit{Valid profit.} In smart contracts, evaluating the profitability of a vulnerability provides a more accurate reflection of its severity~\cite{demystifying,verite}. On one hand, attackers, developers, and auditors are primarily concerned with vulnerabilities that impact the security of funds~\cite{midas,perez2021smart}. On the other hand, a vulnerability may be triggerable but have no practical consequence. For example, in compile versions above 0.8 at Fig.~\ref{fig:motivating_example}, an attacker can trigger Reentrancy without breaking underflow protection as long as the total withdrawn amount is less than \texttt{balances[msg.sender]}. However, in such cases, the attacker gains no profit and is therefore unlikely to exploit in practice.

\vspace{-0.2cm}
\begin{figure}[h]
    \centering
    \includegraphics[width=0.85\textwidth]{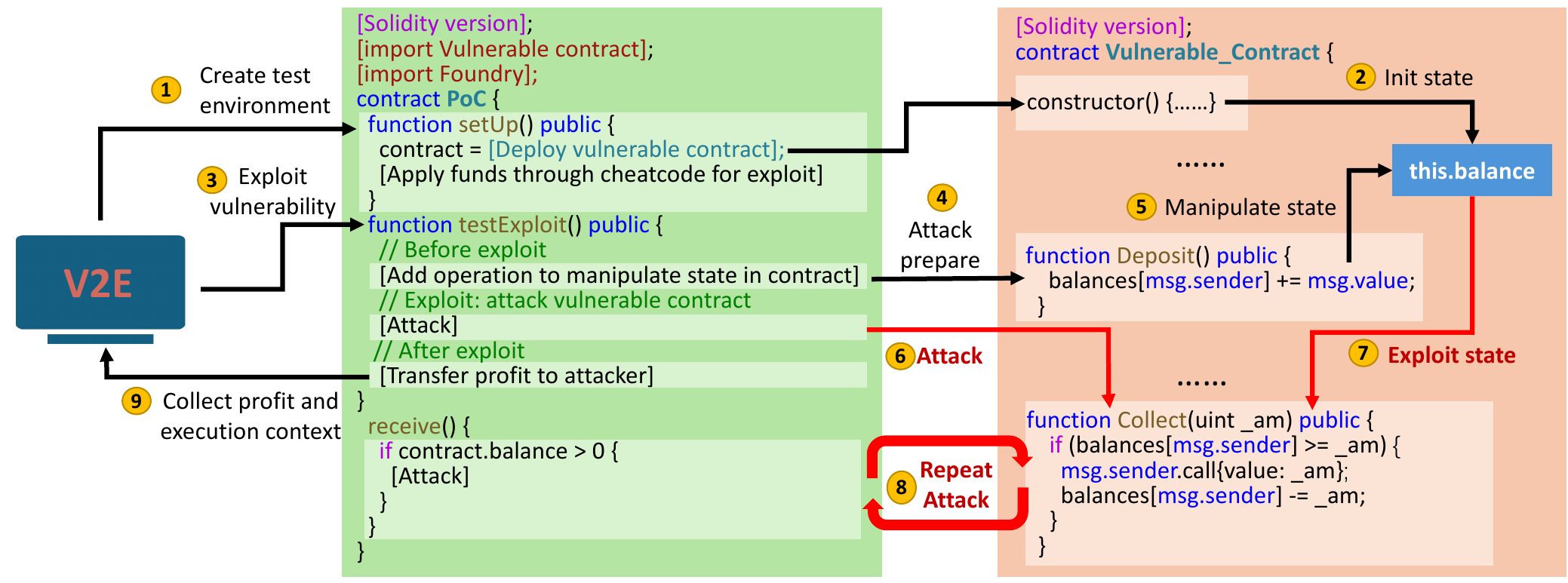}
    \caption{The vulnerability validation process for motivating example in Fig.~\ref{fig:motivating_example} by using PoC.}
    \label{fig:poc_demo}
\end{figure}
\vspace{-0.2cm}

% [Figure]: PoC generation and Execution
\para{Our solution}
To validate the exploitability of a vulnerability, \system{} generates and validates the triggerability and profitability of PoCs based on the Foundry framework~\cite{foundry}. In detail, PoC is a minimal example designed to demonstrate that a vulnerability can be practically exploited. In the context of smart contracts, a PoC is typically a contract that can interact with the target contract. Besides, among PoC development, Foundry is one of the most widely used frameworks, as it provides a complete testing environment and offers a range of cheatcodes. As shown in Fig.~\ref{fig:poc_demo}, \system{} uses a PoC to set up the vulnerability testing environment (steps 1–2), and then proceeds to exploit the vulnerability (step 3). Specifically, it first manipulates critical contract states before the attack (steps 4–5) to simulate realistic conditions. Then, \system{} launches the attack by invoking the target function (steps 6–8). After the attack is completed, \system{} collects profit and execution-related information (step 9) to determine whether the PoC is effective.

% Besides, it can trigger the vulnerable behavior to prove the existence of vulnerability and evaluate the real-world impact.

% As illustrated in Fig.~\ref{fig:poc_demo}, a typical smart contract PoC follows a five-step structure: initialize the testing environment, pre-attack setup, attack, post-attack operations, and evaluate whether the result matches the expected behavior.

We use the example in Fig.~\ref{fig:motivating_example} to illustrate how \system{} identifies the exploitable vulnerability. When \system{} receives the code, it begins by analyzing the vulnerability path and generating a PoC that attempts to exploit Reentrancy. For compile versions below 0.8, \system{} finds that the generated PoC successfully triggers the vulnerability and results in an increase in the attacker’s balance. Therefore, \system{} determines that the vulnerability is exploitable and outputs the result. For versions above 0.8, when \system{} attempts to profit from the vulnerability, it observes that the PoC consistently fails to trigger Reentrancy during execution. In response, it iteratively analyzes the cause of the failure and updates the PoC. However, even when the PoC successfully triggers Reentrancy, \system{} finds that the attacker’s balance does not increase. As a result, it concludes that the vulnerable path is not exploitable and ultimately filters it out.

\subsection{Scope of \system{}}
In this paper, \system{} adopts a stricter standard for vulnerability evaluation. In detail, \system{} considers a detected vulnerability to be valid only if it is both triggerable and profitable, as it can be exploited in the real world. Otherwise, \system{} treats it as a false alarm. 
% Therefore, when a detection tool flags a harmless vulnerability or contract without vulnerability as true

The primary goal of \system{} is to confirm exploitable smart contract vulnerabilities. Specifically, \system{} aims to: (1) validate exploitable vulnerabilities that pose real profits within vulnerable contracts, and (2) eliminate false alarms introduced by existing methods. By doing so, \system{} enables more efficient vulnerability response and remediation under constrained resources. 

Currently, \system{} focuses on five categories defined in the \textit{Smart Contract Weakness Classification~(SWC)}~\cite{swc}: \textit{Unprotected Ether Withdrawal~(UEW), Unprotected Selfdestruct~(US)}, \textit{Reentrancy~(RE)}, \textit{Transaction Order Dependence~(TOD)}, and \textit{Randomness from Chain Attributes~(RCA)}. The reasons are as follows: they are commonly observed SWC from real-world smart contract audits~\cite{dappscan}. Besides, since not all SWC categories represent vulnerabilities with real-world damage, e.g., Code With No Effects, we selectively include only those that can potentially allow an attacker to gain profit when triggered.

\section{Design of \system{}}
\label{sec:designs}
% In this section, we mainly introduce the design choices and the workflow of \system{}.

\subsection{Key Ideas}
\label{subsec:design_choices}
\para{Vulnerability Path Guided PoC Synthesis}
Different vulnerabilities have different exploitation methods. An intuitive approach is to leverage the reasoning capabilities of LLMs to analyze vulnerable contracts and generate PoCs directly. However, PoC generation is a complex coding task, involving vulnerability understanding, exploit path searching, and code synthesis. LLMs lack such strong reasoning abilities~\cite{llmreason}. 

To enable LLMs to generate valid PoCs, \system{} employs fine-grained static analysis, including state dependency and access control analysis, to search for potential vulnerability entry functions. Prior research~\cite{demystifying} shows that over 86\% of exploits (including both inter- and intra-contract vulnerabilities) follow a “state-modify-then-state-read” pattern.
% The insight here is that while vulnerabilities may have different characteristics, exploiting them requires controlling the contract's state. 
Therefore, by analyzing the entry functions that affect the state reading in vulnerable functions, it is possible to build vulnerability exploit paths. Note that smart contracts often include access control mechanisms, such as restricting access to specific accounts or preventing direct external calls. Thus, access control analysis can reduce ineffective detections and false positives. Besides, \system{} initializes the contract in the PoC before testing, enabling support for undeployed contracts. Additionally, to avoid hallucinations caused by excessively long inputs~\cite{xu2024hallucination, li2024dawn, qiu2024longhalqa}, \system{} provides only essential information to the LLM, such as constructors, potential vulnerability paths, and vulnerability features. Then, \system{} guides the LLM to reason step-by-step and output the PoC. We will give more details in Section~\ref{subsec:poc_synthesis}.

\para{Trigger and Profit-Driven PoC Effectiveness Analysis}
Unlike vulnerability detection, using PoC to verify smart contract vulnerability must answer two questions: whether the vulnerability is triggerable and causes real harm. Existing methods are difficult to achieve this purpose. Specifically, static analysis~\cite{defiTainter, etainter} or LLM-based approaches~\cite{gptscan, iaudit} rely on code patterns to identify vulnerabilities, making them unable to handle complex state updates and operation stages in PoCs. Additionally, fuzzing methods~\cite{midas, sFuzz} cannot distinguish state changes caused by cheatcodes, limiting their ability to process intricate interactions in PoC execution. Meanwhile, relying on inline commands~\cite{A1} neither provides fine-grained execution context nor offers extensibility.

As mentioned before, \system{} verifies vulnerability through triggerability and profitability analysis of PoC. In detail, to avoid the influence of cheatcodes on the results, \system{} executes the PoC in a customized EVM and collects the execution context to analyze whether the vulnerability has been triggered. Besides, \system{} records the balance changes during PoC execution and feeds them into LLM for profitability analysis. The key insight here is that in smart contracts, assets can take various forms, including native tokens and other types of tokens, and may involve not only the attacker but also other users. By considering the type and source of each asset, the LLM can accurately determine whether the attacker has gained profit. More details are shown in Section~\ref{subsec:poc_analysis}

% On the one hand, assessing exploitability can help reduce the overall false alarms. On the other hand, profitability helps evaluate the severity of vulnerabilities, as attackers are more likely to exploit vulnerabilities for financial gain. For example, \texttt{sell} in Fig.~\ref{fig:motivating_examples}~(b) theoretically contains Unprotected Ether Withdrawals. However, this vulnerability does not allow the attacker to gain profit~(line 3), so its impact is limited. 

\begin{figure}[t]
    \centering
    \includegraphics[width=0.9\textwidth]{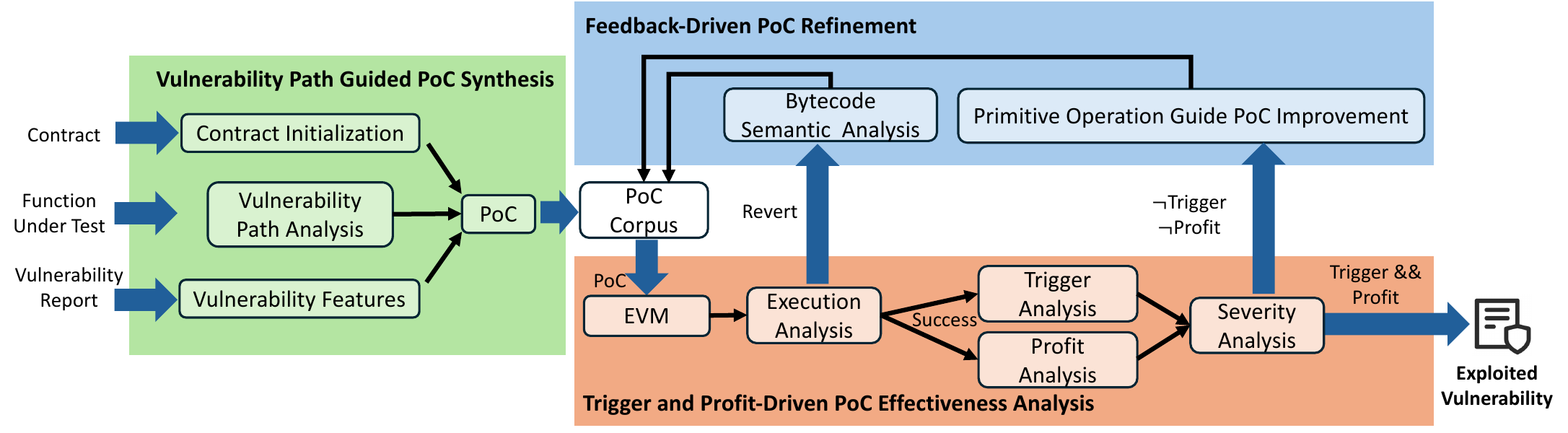}
    \caption{The detailed workflow of \system{}.}
    \label{fig:system_overview}
\end{figure}

\para{Feedback-Driven PoC Refinement}
As mentioned earlier, PoCs generated by LLMs may be invalid, as LLMs cannot account for the impact of dynamic state changes on execution. Therefore, guiding the PoC to explore vulnerability is essential. Since LLM outputs lack diversity~\cite{patchagent}, directly asking LLM to regenerate a PoC is ineffective. Instead, updating the PoC based on execution feedback, such as control flow and data flow changes, is a feasible method and is commonly used in vulnerability detection~\cite{sFuzz, ityFuzz, rethinking}. However, LLM cannot directly utilize this feedback, as smart contracts can only provide bytecode-level feedback during execution.

% To effectively update PoCs, \system{} constructs a PoC test set~(PoC Corpus) and selects a PoC from corpus for execution each time. \system{} updates the PoC's validity based on execution feedback. The updated PoC is then added back to corpus for further testing. 
\system{} feeds the execution context and the failed PoC into LLM for refinement. Although the LLM’s nondeterministic nature may affect how the PoC is modified, it will not affect the performance of \system{}, as Step 2 in \system{} validates the refined PoC and thus ensures its correctness. Specifically, there are two scenarios in which the PoC needs to be adjusted: (1) the PoC execution fails. In this case, \system{} collects execution bytecode and identifies the most critical bytecode, like failed checks, or incorrect state read/write. Next, \system{} uses Source Mapping to recover the corresponding source code from the bytecode. Source Mapping~\cite{source_mapping} is a file generated by the compiler during compilation. It provides detailed records of which source code fragments correspond to each bytecode. Finally, the critical source code and PoC are fed into LLMs for updates; (2) the PoC executes successfully, but the exploitability of the vulnerability cannot be evaluated. Rather than directly modifying parameters, \system{} improves PoC through primitive operations. Primitive operations are a set of predefined PoC update strategies designed based on vulnerability characteristics and stages of PoC execution. The insight here is that different vulnerabilities require different exploitation strategies. For example, in Transaction Order Dependence, the sequence of transactions has a greater impact, while in Randomness from Chain Attributes, finding suitable block properties is more critical. By guiding PoC updates based on the characteristics of the vulnerability, it becomes more likely to confirm the exploitability. More details are shown in Section~\ref{subsec:poc_fine_tuning}.

\vspace{-0.3cm}
\begin{figure}[th]
    \centering
    \begin{tcolorbox}[colframe=black, colback=gray!10, title=PoC Synthesis Prompt Template, fontupper=\small]
        \textbf{System:} You are a smart contract expert, and you need to analyze the smart contract I provide and generate a valid PoC to verify the vulnerability. You must follow the user-defined rules. You can only output one runnable file of PoC without any explanations or markdown. 
        
        \noindent\hdashrule[0.5ex]{\linewidth}{1pt}{3mm}
    
        \noindent\textbf{Contract Initialization:} [Constructor] is the constructor of the contract. You need to analyze the purpose of each parameter and generate appropriate initialization values to deploy the contract.
        
        \noindent\textbf{Vulnerability Path:} [Function] may have {Vulnerability}, and [Related Function] is the function that can modify the state that [Function] depends on. Try to combine these functions and the additional information I provide below to trigger the vulnerability and make profits.
        
        \noindent\textbf{Vulnerability Features:} ......

        \noindent\hdashrule[0.5ex]{\linewidth}{1pt}{3mm}
        
        \noindent\textbf{PoC Requirements:} You need to use forge-std/Test.sol and function setUp to initiate the environment, and exploit vulnerability in function testExploit.

        \noindent\textbf{Additional Information:} e.g., constant states read by [Function], block number to be tested......

        \noindent\hdashrule[0.5ex]{\linewidth}{1pt}{3mm}

        [Function]

        [Constructor]

    \end{tcolorbox}
  \caption{The prompt template used for PoC synthesis.}
  \label{fig:poc_synthesis}
\end{figure}
\vspace{-0.5cm}

% Therefore, based on the analysis of vulnerability, \system{} defines a set of operations to help explore PoC.

\subsection{Workflow of \system{}}
Fig.~\ref{fig:system_overview} shows the detailed workflow of \system{}. \system{} takes contract, function under test, and vulnerability report as input, automatically verifies the vulnerability through PoC, and outputs exploited vulnerability.  Specifically, in Vulnerability Path Guided PoC Synthesis, \system{} uses fine-grained static analysis to identify potential vulnerability entry functions in the contract and constructs vulnerability paths based on these functions. Then, \system{} incorporates contract initialization requirements, vulnerability paths, and vulnerability characteristics into the prompt, guiding the LLM to generate PoCs. After that, these PoCs are stored in PoC Corpus for further analysis. In Trigger and Profit-Driven PoC Effectiveness Analysis, \system{} retrieves one PoC at a time from the corpus for execution and collects execution context to evaluate its validity. When \system{} identifies a PoC that can both trigger vulnerability and make profits, it outputs the PoC and confirms the vulnerability. Otherwise, \system{} employs PoC upgrade. In Feedback-Driven PoC Refinement, if execution fails, \system{} extracts and analyzes critical bytecode. Simultaneously, if no evidence is found, \system{} improves the PoC by using primitive operations and adds the updated PoC into the corpus for further analysis.

% \para{Our solution}
% Again, take Fig.~\ref{fig:motivating_examples} as examples. For function \texttt{withdraw} in Fig.~\ref{fig:motivating_examples}~(a), \system{} identifies that \texttt{deposit} at line 8 is a potential vulnerability entry point and constructs PoCs. However, due to the protection on lines 2, the PoC consistently fails the exploitability check. Eventually, \system{} concludes that this is a harmless vulnerability. For function \texttt{sell} in Fig.~\ref{fig:motivating_examples}~(b), \system{} identifies \texttt{buy} at line 7 as a potential entry point and generates PoCs. However, \system{} finds that the generated PoC cannot yield any profit and marks this vulnerability as harmless. For function \texttt{solvePuzzle} in Fig.~\ref{fig:motivating_examples}~(c), \system{} identifies \texttt{solvePuzzle} itself as the entry point based on the characteristics of Transaction Order Dependence, and the generated PoC passes both exploitability and profitability checks. Then, \system{} confirms the vulnerability and outputs the valid PoC.

\section{Approach Details}
\label{sec:approach_details}
% In this section, we present detailed implementations of \system{}.

\subsection{Vulnerability Path Guided PoC Synthesis}
\label{subsec:poc_synthesis}
To generate specific PoCs, \system{} employs fine-grained static analysis to analyze potential vulnerability paths and includes detailed deployment requirements along with vulnerability features in the prompt, guiding the LLM to reason step-by-step and produce PoCs.

\para{Vulnerability path analysis}
For a given function under test, \system{} uses the following formula to find potential entry functions:
\begin{align}
    F_{dep}(f_{test}) = \{f \in C | Write(f) \cap Read(f_{test})\neq \emptyset \& \neg Access(f) \} 
\end{align}
where $f_{test}$ represents the function under test, $F_{dep}$ denotes the set of entry functions for $f_{test}$, $C$ refers to the contract under test, $Write$ represents the set of states updated by function, $Read$ signifies the set of states read by the function, $Access$ is the access control mechanism in $f$. Specifically, this formula is composed of two components: state read-write dependency analysis and access control analysis. State read-write dependency analysis first identifies the states that the tested function depends on, and then searches the contract for entry functions that can modify those states. After that, access control analysis determines whether the function has access control by primarily analyzing the presence of modifiers like \texttt{onlyOwner}, as modifiers are widely used to implement access control in smart contracts (over 66\% of contracts~\cite{fang2023beyond}). Moreover, although contracts may implement access control through other mechanisms (e.g., \texttt{require}), resulting in incorrect PoCs, the PoC effectiveness analysis in Section~\ref{subsec:poc_analysis} can substantially mitigate such false positives.

% \begin{figure}[tbp]
%     \centering
%     \includegraphics[width=0.9\textwidth]{figure/PoC_generation_prompt.pdf}
%     \caption{Prompt template for PoC Synthesis.}
%     \label{fig:poc_synthesis}
% \end{figure}

\para{PoC synthesis}
For each $f$ in $F_{dep}(f_{test})$, \system{} attempts to combine $f_{test}$ and $f$ to construct a potential exploit path and generate the PoC. In detail, \system{} employs the prompt template shown in Fig.~\ref{fig:poc_synthesis}. This template is composed of three sections. It begins with a system prompt that provides an overview of the task and sets the LLM's role. 
The middle part is the user prompt, which includes three steps for PoC generation. For subsequent vulnerability testing, \system{} instructs the LLM to first deploy the contract based on constructor parameters. Then, it generates the PoC according to the vulnerability path and specific vulnerability characteristics.
The final part of the prompt provides the constraints for PoC synthesis. Specifically, \system{} utilizes Foundry for PoC execution and evaluation. To standardize the testing process, \system{} mandates the use of \texttt{setUp} for test initialization and \texttt{testExploit} as the entry point for the exploit. Besides, to adapt different requirements for various vulnerabilities and contracts, \system{} also injects additional information, like contract invariants or block number, and instructs the LLM on its proper utilization.

\system{} stores all generated PoCs in the PoC Corpus and continuously retrieves PoCs from the corpus for execution, evaluating and updating them based on their triggerability and profitability.

\subsection{Trigger and Profit-Driven PoC Effectiveness Analysis}
\label{subsec:poc_analysis}

\system{} mainly performs PoC effectiveness analysis in this step. Particularly, \system{} feeds the PoC into a customized EVM for execution. First, the EVM invokes function \texttt{setUp} in the PoC to initialize the execution environment, such as deploying the vulnerable contract and allocating sufficient funds. \system{} ignores all actions performed within \texttt{setUp} to ensure the accuracy of the PoC evaluation. Then, \system{} invokes function \texttt{testExploit} in the PoC to perform the attack and collects the execution context during this process. Note that in this step, \system{} ignores all control-flow and data-flow changes caused by cheatcode operations. Finally, \system{} checks step-by-step whether the PoC executes successfully, triggers the vulnerability, and generates profit. If any step yields a negative result, \system{} terminates the analysis and proceeds to update the PoC with strategies in Section~\ref{subsec:poc_fine_tuning}.

\para{Execution analysis}
After PoC execution, \system{} analyzes whether the final opcode executed by the PoC is \texttt{revert}, as the EVM returns opcode \texttt{revert} when execution fails. Therefore, if it is \texttt{revert}, the PoC is marked as a failure. Otherwise, execution is deemed successful.

\vspace{-0.1cm}
\input{tables/vulnerability_trigger}
\vspace{-0.2cm}

\para{Trigger analysis}
To determine whether PoC triggers a vulnerability during execution, \system{} defines different trigger rules for each vulnerability. The detailed rules are shown in Table~\ref{tab:detection_rules}. 
\begin{itemize}
    \item \textbf{Unprotected Ether Withdrawal}. \system{} checks whether the PoC contains any $Transfer$ to the transaction sender $tx.origin$.

    \item \textbf{Unprotected Selfdestruct}. \system{} determines whether the PoC invokes opcode $Selfdestruct$.
    
    \item \textbf{Reentrancy}. \system{} analyzes whether two calls $c_i$ and $c_j$ occur in execution, where $c_j$ is a nested call inside $c_i$ and they invoke the same function.

    \item \textbf{Transaction Order Dependence}. \system{} examines whether the PoC contains two calls, $c_i$ and $c_j$, such that $c_i$ updates a specific storage $slot$ and $c_j$ subsequently reads $slot$'s value.

    \item \textbf{Randomness from Chain Attributes}. \system{} identifies whether there is an opcode $op$ related to reading block attributes, like block number, timestamp, and block hash.
\end{itemize}

When the corresponding vulnerability rule is observed during PoC execution, \system{} considers the PoC to have successfully triggered the vulnerability.

\para{Profit analysis}
\system{} evaluates the attacker’s profit to assess the severity of the vulnerability. Specifically, \system{} instructs the PoC to output the attacker’s balance changes before and after the exploit via EVM events. Since blockchains contain native assets such as Ether, as well as various types of tokens, such as ERC-20 tokens and non-fungible tokens~(NFTs), these balance changes are then fed into LLM to determine whether the exploit is financially profitable. 
% Besides, if the PoC is deemed profitable, \system{} considers it valid and outputs the PoC accordingly.

\vspace{-0.1cm}
\input{algorithm/execution_fail_analysis}

\para{Severity analysis}
\system{} terminates execution either when a valid PoC is found or when a predefined execution timeout is reached. Upon termination, \system{} analyzes the execution results of all PoCs in the corpus and provides one of the following three conclusions:

1.	\textbf{Exploitable}: If \system{} identifies a PoC that both successfully triggers the vulnerability and yields profit for attackers, it outputs $Exploitable$, indicating that the vulnerability poses a real threat.
\begin{align}
    \exists PoC \in corpus, Trigger(PoC)\&\&Profit(PoC) => Exploitable
\end{align}

2.	\textbf{Non-exploitable}: If, by the end of validation, \system{} finds that there is a PoC that can either trigger the vulnerability without profit or generate profit without triggering the vulnerability, it considers the vulnerability to be of low severity or nonexistence, and outputs $Non-exploitable$.
\begin{align}
\begin{split}
    \exists PoC \in corpus, Trigger(PoC)\&\& \neg Profit(PoC) || \\
    \neg Trigger(PoC)\&\& Profit(PoC) => Non-exploitable
\end{split}
\end{align}

3.	\textbf{Manually Check}: If no PoC is found that can either trigger the vulnerability or generate profit, \system{} outputs $Manually \ Check$, suggesting that the vulnerability requires further exploration.
\begin{align}
    \forall PoC \in corpus, \neg Trigger(PoC)\&\& \neg Profit(PoC)=> Manually \ Check
\end{align}

\subsection{Feedback-Driven PoC Refinement}
\label{subsec:poc_fine_tuning}

When the PoC fails verification, \system{} performs PoC optimization based on the execution feedback through bytecode semantic analysis and primitive operation.

\para{Bytecode semantic-based PoC update}
If a PoC fails, \system{} uses the algorithm shown in Algorithm~\ref{algor:execution_fail} to update the PoC. Specifically, \system{} collects all the bytecode from the execution and analyzes it to extract the execution failure output. Then, \system{} analyzes bytecode in reverse order to find the critical point of failure. To confirm the semantics of the bytecode, \system{} first compiles the contract related to the bytecode and retrieves the source mapping file. \system{} then locates the source code fragment corresponding to the bytecode through this file. When finding a bytecode whose related source code is a check (e.g., \texttt{require} or \texttt{assert}) or a state-update statement, \system{} identifies this as the critical failure statement. \system{} focuses on these two types of statements because contract execution failures are typically caused by failing checks or state-update errors, such as transferring with insufficient balance. Finally, \system{} provides the failing statement, its corresponding function, and the failure reason output by the EVM to the LLM for the PoC update. The updated PoC is added to the corpus for further in-depth analysis.

\input{tables/primitive_operation}
\para{Primitive operation guide PoC improvement}
A PoC may execute without errors but still fail to validate a vulnerability, for instance, by not confirming its triggerability or profitability. In such cases, \system{} optimizes the PoC by selecting appropriate operations from a predefined set, based on the vulnerability type and the specific stage of failure. The specific names and rules of primitive operations are shown in Table~\ref{tab:primitive_operation}.

\begin{itemize}
    \item \textbf{add\_user}. This operation is designed to enhance the profitability of RCA, TOD, and UEW. In detail, these vulnerabilities are all state-sensitive. However, after the PoC is deployed, the contract is in initial state, and the state-modifying operations provided by PoC may be insufficient. Thus, \system{} simulates state changes by adding additional user operations.

    \item \textbf{change\_invoker}. This operation is used to aid in exploring RE, UEW, and US. These vulnerabilities may impose constraints on the caller. For example, the vulnerable path may only be triggered when invoked by the deployer. To address this, \system{} attempts to explore different execution paths by varying the function caller.

    \item \textbf{change\_order}. This operation is designed to improve the profitability of TOD and UEW, which are sensitive to the order of function calls. \system{} explores profitability by performing execution reordering to trigger different sequences of state modifications.
    % For example, if a function can be invoked by both attackers and the victim. If the attacker calls it first, they can profit, but if the call comes after the victim, the exploit fails. Thus, \system{} rearranges the order of function calls within the PoC to examine profitability.

    \item \textbf{modify\_block}. This operation is used to explore the triggerability of RCA. Since RCA relies on block attributes, such as the block number and timestamp, to simulate randomness, \system{} attempts to trigger different execution paths by modifying current block properties.

    \item \textbf{change\_argument}. This operation applies to all types of vulnerabilities and all stages of analysis. By modifying function parameters and the amount of Ether sent with function calls, \system{} attempts to trigger alternative execution paths and explore how different state conditions impact the profitability.
\end{itemize}

For all selected suitable primitive operations, \system{} sequentially applies them to modify the PoC. All newly generated PoCs are then added back into the corpus for further analysis.

\section{Evaluation}
\label{sec:evaluation}
In this section, we first provide the datasets used in the evaluation and introduce our evaluation setup. Then, we show the evaluation results of \system{}.

\input{tables/dataset}
\vspace{-0.3cm}

\subsection{Implementation and Evaluation Setup}
\label{subsec:implementaion}
\para{Dataset}
We collect 264 vulnerable contracts from SmartBugs~\cite{smartbugs}, as it provides a large number of open-source and peer-reviewed contracts. These contracts are divided into $D_{manual}$ and $D_{onchain}$ based on their origin. Specifically, $D_{manual}$ contains 64 manually constructed vulnerable smart contracts, while $D_{onchain}$ consists of 200 real-world contracts deployed on-chain. Note that SmartBugs contains vulnerabilities that are out of our scope, such as Unchecked Low-Level Calls, as well as some contracts that are incompatible with Foundry~\cite{foundry}. Therefore, we apply the following procedures to construct both datasets. The distribution of vulnerabilities is shown in Table~\ref{tab:dataset}.

% We use the curated dataset~($D_{manual}$) and consolidated ground truth dataset~($D_{onchain}$) provided by SmartBugs~\cite{smartbugs} to evaluate \system{}, as they offer a large collection of open-source and peer-reviewed vulnerable contracts. In detail, we select 64 manually crafted vulnerable contracts from $D_{manual}$. Besides, all contracts~(a total of 200) in $D_{onchain}$ are vulnerable contracts deployed on-chain. Note that these datasets contain vulnerabilities that are out of our scope, such as Unchecked Low-Level Calls, as well as some contracts that are incompatible with Foundry~\cite{foundry}. Therefore, we apply the following procedures to refine both datasets. The distribution of vulnerabilities is shown in Table~\ref{tab:dataset}.

$\bullet$ \textbf{Filtering vulnerabilities.} As mentioned earlier, \system{} refers to the common SWC categories summarized by real-world audits~\cite{dappscan} and selects five types of vulnerabilities based on their potential for profit. Since $D_{manual}$ does not provide SWC labels, we manually craft PoCs, collect execution traces (i.e., paths) and fine-grained state changes during PoC execution. Then, we leverage the SWC features to select 64 vulnerable contracts. For $D_{onchain}$, we perform filtering based on the SWC tags included in the dataset. Initially, $D_{onchain}$ contains 761 vulnerable contracts. Due to time constraints, we choose a sample size of 200 based on a 90\% confidence level and a 5\% margin of error. We then determine the number of vulnerabilities in each category according to their distribution and conduct sampling accordingly.
% we select a representative subset of 200 contracts.
% we manually classify these contracts based on the vulnerability information provided. 

$\bullet$ \textbf{Filtering out contracts with complex constructor parameters.} When contract deployment depends on complex parameters, such as addresses or bytes, \system{} may not know what values to assign, leading to test failures. Therefore, contracts with such deployment requirements are excluded from evaluation. However, it is worth noting that this procedure does not affect the representativeness of the dataset. Specifically, we inspect 31 contracts excluded during sampling and find that their vulnerability patterns are fully covered by remaining cases, and 87\% of complex parameters are irrelevant to vulnerable locations.
    
$\bullet$ \textbf{Updating contract version to above 0.6.} To satisfy the minimum Solidity version supported by Foundry (i.e., 0.6)~\cite{forge_std}, we manually upgrade all contracts written in versions earlier than 0.6. Note that this upgrade is restricted to syntax-level compatibility adaptations required for compilation, such as updating deprecated keywords or language constructs. The whole process does not involve any intentional modification of the contract logic, control flow, or behavior. To validate that these changes do not alter contract semantics, we construct a PoC for each contract that requires an upgrade. Then, we check whether the PoC maintains consistent results before and after update. The results show that no vulnerabilities exhibit different behaviors due to the update.
% Foundry only supports Solidity versions greater than 0.6~\cite{forge_std}. Thus, for contracts in the dataset with lower versions, we manually upgrade their version to 0.6, and resolve any compilation errors. To ensure new version does not alter results, we construct a PoC for each contract that requires an upgrade. Then, we check whether the PoC maintains consistent results before and after the update. The results show that no vulnerabilities exhibit different behaviors due to the update.
% Note that the new features introduced in version 0.6 have no impact on the vulnerability. Besides, this process does not modify the execution logic. Therefore, vulnerabilities in the contract will persist.

$\bullet$ \textbf{Labeling the severity of vulnerability.} SmartBugs only provides vulnerability labels but does not include information on the exploitability, like whether they are triggerable and profitable. Therefore, we invite two experts with at least two years of experience in smart contract auditing to assess 264 vulnerabilities independently. Specifically, if an expert determines that a vulnerability could be exploited, they are required to write a PoC. When the two experts reach a consensus, we label exploitable vulnerabilities as Exploitable and others as Non-Exploitable. In cases where the two experts disagreed, a third expert is brought in to validate the PoC and provide a final decision.

\para{Implementation}
\system{} leverages the Foundry-based execution engine provided by ityFuzz and further builds customized PoC execution and fine-grained state collection logic on top of it, enabling PoC validity analysis. Besides, \system{} implements the PoC synthesis and refinement modules from scratch. All experiments in our evaluation are conducted on a machine with two Intel(R) Xeon(R) Gold 5218R CPU @ 2.10GHz, 512GB RAM, and Ubuntu 20.04.4 OS. During the experiments, we use three models: GPT-4o (2024-11-20)~\cite{gpt}, DeepSeek R1 (2025-05-28)~\cite{guo2025deepseek}, and Claude Sonnet 4 (2025-05-14)~\cite{claude_sonnet_4}. By default, \system{} primarily uses GPT-4o in the experiments. Regarding model parameters, for PoC generation and refinement, we set the temperature to 1 to encourage the model to produce more diverse PoCs. In all other stages, we set the temperature to 0 to improve determinism and stability. For the remaining parameters, we follow the configuration used in GPTScan~\cite{gptscan}. In our preliminary experiments, we observe that over 70\% of vulnerabilities can be confirmed within 10 minutes. Moreover, longer execution times increase LLM token overhead without providing benefits. Thus, we set a time limit of 30 minutes for trade-offs. When the timeout is reached, \system{} terminates the analysis and produces a final conclusion based on the results of all PoCs. Besides, we find that providing prior context biases the model toward producing responses that are similar to earlier outputs, regardless of whether those previous responses are correct. Therefore, \system{} clears context before each LLM request.
% Parameters in models are set to default values. \system{} clears context before each LLM request. Additionally, for each vulnerability, we set a time limit of 30 minutes. 

\para{Evaluation metrics}
In this paper, we use TP~(True Positive) to denote correctly identifying an exploitable vulnerability, FP~(False Positive) to denote incorrectly labeling a non-exploitable vulnerability or contract without vulnerability as exploitable, TN~(True Negative) to denote correctly identifying a non-exploitable vulnerability or contract without vulnerability, and FN~(False Negative) to denote misclassifying an exploitable vulnerability as non-exploitable.

\para{Research questions}
We primarily focus on the following four research questions:
\begin{enumerate}[label=\textbf{RQ\arabic*.},leftmargin=1.4cm]
    \item How effective is \system{} in validating exploitable vulnerability?
    \item How effective is \system{} compared with existing methods?
    \item How effective is \system{} in eliminating false alarms?
    \item What is the impact of each module in \system{} on verifying vulnerabilities?
    \item What is the performance overhead of \system{}?
\end{enumerate}

\vspace{-0.2cm}
\input{tables/profitable_analysis}
\vspace{-0.3cm}

\subsection{Effectiveness of \system{} in Validating Exploitable Vulnerability}
\label{subsec:verifying_vulnerabilites}
To answer RQ1, we evaluate \system{} with $D_{manual}$ and $D_{onchain}$. Specifically, for each PoC that \system{} deems exploitable, we let two experts independently audit it. If both experts agree that the PoC successfully triggers the vulnerability and enables the attacker to profit, we consider it a true positive identified by \system{}. In cases of disagreement, a third expert is invited to review the PoC and provide the final judgment. Table~\ref{tab:verifying_vulnerability} shows the final results. As shown in the table, \system{} successfully confirms 102 out of 124 exploitable vulnerabilities, achieving a precision of 91.9\% and a recall of 82.3\%. Meanwhile, it correctly identified 71 out of 140 non-exploitable vulnerabilities, accounting for 50.7\%. Note that 4 FPs, 1 FN, and 3 exploitable MCs and 12 non-exploitable MCs are caused by LLM analysis error. We will provide a further analysis of these cases in Section~\ref{subsec:ablation_study}.

\para{False Positives}
Through the analysis of false positives, we identify the following main reasons: (1) Lack of specific token prices~(2 FPs). Since token prices are determined by the markets and change over time, \system{} does not know the price of tokens (e.g., ERC20 tokens) in the contract and defaults to a 1:1 exchange rate with native tokens, like Ether. For example, if the attacker spends 1 ether to obtain 100 tokens, \system{} considers the attacker to have made a profit of 99 tokens. However, this approach may lead to incorrect estimations of potential attack profits, resulting in false positives. (2) Bypassing checks that should not be bypassed~(3 FPs). When execution fails due to permission checks, such as requiring the caller to be the contract deployer, \system{} may attempt to directly let the attacker deploy the contract and bypass the check.

\begin{figure}[h]
    \centering
    \includegraphics[width=0.45\textwidth]{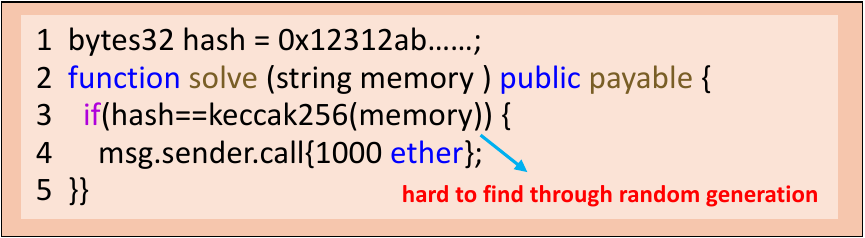}
    \caption{A False Negative caused by a lack of mathematical reasoning capabilities.}
    \label{fig:false_negative}
\end{figure}
\vspace{-0.2cm}

\para{False Negatives and Manually Check}
The causes of false negatives and exploitable Manually Check are similar, and can be attributed to the following reasons: (1)~\system{} fails to bypass \texttt{tx.origin}~(2 FNs and 5 exploitable MCs). All transactions in the PoC are executed as internal transactions, meaning the caller is a fixed externally owned account. However, some contracts enforce \texttt{tx.origin} checks to ensure that only authorized users can trigger certain functions. As a result, the PoC repeatedly fails the check and cannot proceed with the exploit. (2)~\system{} lacks complex mathematical capabilities~(4 FNs and 7 exploitable MCs). Due to the limited arithmetic reasoning ability of LLMs, \system{} struggles to infer precise inputs when the contract involves complex computations. For example, as illustrated in Fig.~\ref{fig:false_negative}, there exists a profitable TOD in function \texttt{solve}. In a real-world scenario, once a user discovers a valid input and sends a transaction, the attacker can frontrun the transaction by replicating the input. Although \system{} successfully generates the correct transaction sequence structure, it can only attempt to trigger the vulnerability by randomly generating candidate inputs, lacking the ability to compute them directly.

For non-exploitable Manually Check, the primary reason lies in \system{}’s conservative strategy~(48 non-exploitable MCs). Specifically, to reduce the impact of LLM uncertainty, \system{} only proceeds to assess vulnerability when the PoC passes either the triggerability or profitability analysis. However, some vulnerabilities are merely theoretical and fail during actual execution. In such cases, \system{} cannot perform trigger and profit analysis, and ultimately leads to Manually Check.
% For example, if PoCs generated by \system{} consistently fail to execute in the case of \texttt{Collect} shown in Fig.~\ref{fig:motivating_examples}(a), attempting to trigger Reentrancy will revert due to the check on line 2. As a result, the PoCs generated by \system{} consistently fail to execute, preventing further analysis, and ultimately leading to Not Sure.

\vspace{-0.2cm}
\input{tables/baseline}
\vspace{-0.2cm}

\subsection{Comparison with Existing Methods}
\para{Comparison with A1}
As mentioned earlier, the most relevant tool to \system{} is A1, a SOTA framework generating executable PoCs for smart contract vulnerability detection~\cite{A1}. However, since the original A1 only provides a demo version~\cite{a1_demo} to the public, we cannot perform an apples-to-apples comparison. As our best effort, based on the framework in~\cite{a1_demo}, we reimplement A1 step by step, following the procedure and its core prompts described in the A1 paper. We refer to this re-implementation version as $A1_{re}$. 
To ensure the reliability of $A1_{re}$, we evaluate it using the original A1’s benchmark (36 incidents). Note that, as our best effort, we follow the same parameters specified in A1. For example, we aggregate the outputs of the same six models to derive the final answer, running two trials with five rounds per iteration. In the end, $A1_{re}$ identifies 22 incidents, whereas A1 identifies 26. These results indicate that $A1_{re}$ is relatively comparable to A1. Finally, we apply $A1_{re}$ to $D_{manual}$ and $D_{onchain}$, and compare the results with \system{}. As shown in Table~\ref{tab:baseline}, \system{} significantly performs better than $A1_{re}$. Particularly, $A1_{re}$ incurs much higher FPs and FNs than \system{} on the two datasets.
The high efficacy of \system{} is mainly attributed to the adoption of off-chain testing and our customized PoC executing environment. In this way, \system{} can collect fine-grained context information and explore a larger state space, thereby validating more exploitable vulnerabilities.

% A1 releases the benchmark (36 incidents) it used. Thus, in this part, we evaluate \system{} on this dataset by discussing their individual performance.
% %
% Specifically, we identify the root causes of the reported attacks based on public disclosures and then filter out 11 vulnerabilities that \system{} can cover (including 1 RE, 1 RCA, and 9 UEW). We conduct two rounds of experiments. First, since detailed attack processes (e.g., PoCs~\cite{a1_attack_1, a1_attack_2} and postmortem~\cite{a1_attack_3}) for these 11 cases are all publicly available online, we follow A1 and provide only the function labels to \system{} to examine whether the model memorizes these attacks. This setting helps reduce potential bias introduced by memorization. The results show that \system{} fails to verify any vulnerability, and memorization has no effect. Next, we provide the complete information to \system{} and re-run the evaluation. \system{} correctly identifies all vulnerabilities, with 100\% precision and 100\% recall. In comparison, A1 identifies 9 exploitable vulnerabilities with 100\% precision and 81.8\% recall.

\para{Comparison with multi-agent and static-based methods}
To better demonstrate the effectiveness of \system{}, we draw inspiration from iAudit~\cite{iaudit} and use a framework combined with multi-agent and static analysis, called $LLM_{multi}$, to analyze vulnerabilities in both $D_{manual}$ and $D_{onchain}$. Specifically, we first generate PoCs based on vulnerability path guided PoC synthesis for the given vulnerability. Then, an LLM analyzes whether the PoC successfully triggers the vulnerability and yields profit, providing a detailed justification. Finally, another LLM evaluates the reasoning. If the justification is deemed valid, a conclusion is produced. Otherwise, the process loops back to reassess and regenerate the explanation. The detailed results are shown in Table~\ref{tab:baseline}. In detail, LLMs are unable to execute code and obtain instantaneous runtime state values. As a result, they assess vulnerability exploitability solely based on static code snippets. This limitation leads to inaccurate judgments and introduces a large number of false positives and false negatives.

\vspace{-0.2cm}
\input{tables/false_alarms}
\vspace{-0.3cm}

\subsection{Effectiveness of \system{} in Eliminating False Alarms}
\label{subsec:reducing_false_arlarms}
To answer RQ3, we use two actively maintained static analysis tools, Slither~\cite{slither} and Mythril~\cite{mythril}, as well as one fuzzing tool, Confuzzius~\cite{confuzzius} to identify vulnerabilities in both $D_{manual}$ and $D_{onchain}$. We invite two experts to independently validate the detection results produced by the three tools. If they agree that a reported issue is a true vulnerability, they write a PoC. If there is any disagreement between the two experts, a third expert will assess the PoC validity, and the three experts together will reach a final decision. Note that we do not perform a comprehensive recall analysis by aggregating the results of the three tools, as \system{} mainly aims to validate whether reported vulnerabilities are exploitable, rather than detecting vulnerabilities. In practice, if a detection tool misses a vulnerability, \system{} cannot conduct further analysis. In addition, although \system{} may misclassify real vulnerabilities as FNs, this rarely occurs (with recall of 82.3\% in our experiments). Besides, we exclude other testing tools such as ityFuzz~\cite{ityFuzz}, Verite~\cite{verite}, or Sailfish~\cite{sailfish}, as they do not cover all vulnerabilities \system{} targets. 

We then use \system{} to analyze the detection results reported by these tools. We let two experts review the outcomes of \system{}. If \system{}’s detection result is inconsistent with the ground truth, we label it as either a false negative or a false positive. As shown in Table~\ref{tab:false_alarm}, although the detection tools report a large number of vulnerability warnings, most of them are false positives. In contrast, by analyzing both triggerability and profitability, \system{} successfully assesses the majority of these vulnerabilities, effectively reducing false alarms.

\subsection{Effectiveness of Each Module in \system{}}
\label{subsec:ablation_study}
To answer RQ4, we use $D_{manual}$ and $D_{onchain}$ to analyze the impact of different LLMs and individual components on \system{} overall effectiveness.

\vspace{-0.15cm}
\input{tables/different_llm}
\vspace{-0.1cm}

\para{Impact of different LLMs}
We evaluate the performance of \system{} under GPT-4o, DeepSeek R1, and Claude Sonnet 4, respectively. The results are shown in Table~\ref{tab:different_llms}. Since PoC generation and update involve understanding the code and the vulnerability, and generating new code, SOTA LLMs perform differently. Specifically, we find that GPT-4o identifies the highest number of true positives and has the fewest false negatives. DeepSeek R1 produces the lowest number of false positives, while Claude Sonnet 4 results in the fewest cases requiring manual confirmation. Overall, GPT-4o achieves the best precision and recall among the three models. Through the analysis, we find that the primary cause of LLM misjudgments is the lack of sufficient source code context. This is because \system{} explores a single entry function for each PoC. However, when a vulnerability requires manipulation of multiple functions, the PoC's effectiveness is compromised because LLM lacks context to invoke. This design is a deliberate trade-off, as PoC synthesis is a complex task, and minimizing the LLM's reasoning load is crucial for improving generation outcomes. 
% On the other hand, some functions achieve their logic, such as access control, by invoking other functions. \system{} does not explicitly provide these function implementations and instead relies on the LLM to infer their behavior from function names. As a result, this may lead to incorrect analysis. This trade-off is made to control input length, which helps reduce hallucination and token consumption.

\input{tables/ablation_study}

\para{Effectiveness of vulnerability path analysis}
To verify the impact of vulnerability path analysis on PoC generation, we remove it. Instead, we provide the LLM with all function signatures in the contract and instruct it to generate an exploit path based on its understanding of the vulnerability. The result is shown in Table~\ref{tab:ablation_study}. Although the vulnerability data in SmartBugs has been used to train LLMs, it is still difficult for LLMs to generate effective PoC. This highlights the necessity of providing LLMs with sufficient information (i.e., vulnerability path and potentially vulnerable functions). Moreover, excessively long inputs increase the likelihood of hallucinations. As a result, it fails to generate valid PoCs in many cases, leading to many vulnerabilities remaining unconfirmed.

\para{Effectiveness of trigger rule-based analysis}
We separately investigate the effectiveness of trigger and profit analysis. To validate the trigger rules, we ask the LLM to directly judge the PoC. Specifically, if LLM considers the PoC valid, the PoC is marked as triggerable. Otherwise, we select corresponding primitive operations to update PoC based on the reason and the type of vulnerability. The results are shown in Table~\ref{tab:ablation_study}. Since LLM does not execute code, it tends to overlook how runtime states affect vulnerability triggerability, resulting in more FPs and FNs.

\para{Effectiveness of LLM-based profit analysis}
To validate the effectiveness of profitability analysis, we design a set of rules to determine whether an attack yields profit. Specifically, we collect all native-token transfers (e.g., ETH) and token transfer logs (e.g., ERC-20 Transfer events) produced during execution to analyze the attacker’s profit. As shown in Table~\ref{tab:ablation_study}, due to the heterogeneity of assets in smart contracts, rule-based profit checking may miss certain value-transfer signals, leading to more false negatives.

% To evaluate the effectiveness of trigger and profit-driven analysis in assessing PoC validity, we replace the analysis with an LLM-based approach. Specifically, we instruct LLM to determine the validity of the generated PoC. If LLM considers the PoC valid, the analysis finishes. If the LLM concludes that the PoC either fails to trigger the vulnerability or fails to generate profit, we select corresponding primitive operations to update PoC based on the reason and the type of vulnerability. The results are shown in Table~\ref{tab:ablation_study}. Since the LLM does not execute code, it fails to account for how state changes affect the exploitability of a vulnerability when analyzing PoC validity. As a result, it produces lots of false positives and false negatives. Moreover, most of them cannot be executed successfully.

\para{Effectiveness of feedback-driven PoC refinement}
To evaluate the effectiveness of feedback-guided PoC exploration, we remove this component from the pipeline. Specifically, when a PoC requires updating, \system{} directly feeds the original PoC and its execution results into the LLM and prompts it to generate a new PoC. As shown in Table~\ref{tab:ablation_study}, due to the lack of fine-grained execution feedback, LLM cannot identify the specific reasons why a PoC is invalid. Moreover, consistent with observations in prior studies~\cite{fan2023automated, xia2024automated}, we find that, without additional constraints, LLMs tend to make only local edits, such as changing parameter values, rather than revising the code structure, regardless of the temperature setting. As a result, the generated PoCs are overly similar, which limits their ability to explore vulnerability.
% Moreover, in the absence of execution feedback, the PoCs generated by the LLM tend to become increasingly similar, which limits its ability to explore the vulnerability.

\subsection{Overhead of \system{}}
\label{subsec:overhead}
\system{} takes an average of 17.5 minutes to analyze a vulnerability. The most time-consuming phase is PoC refinement, averaging 9.1 minutes. This indicates that PoCs generated by LLM are often not immediately suitable for validating vulnerability and require continuous iteration by \system{}. The second-largest time cost is PoC compilation, with an average cost of 6.3 minutes. This is because \system{} needs to compile both the initially generated and all subsequently updated PoCs during the validation process. In contrast, since \system{} is an off-chain testing tool, PoC execution is very fast, taking an average of 1.8 minutes. Therefore, the time overhead of \system{} is acceptable.

We record the token usage for each API request, including both input and output, and calculate the total cost based on OpenAI’s official pricing~\cite{gpt_price}. The total cost of running \system{} is 187.4 USD, averaging 0.71 USD per vulnerability. Consistent with the time cost analysis, PoC update incurs the highest token usage, with an average cost of 0.53 USD. Since \system{} is able to provide only essential information to the LLM, the overall cost remains within an acceptable range.

\begin{figure}[h]
    \centering
    \includegraphics[width=0.8\textwidth]{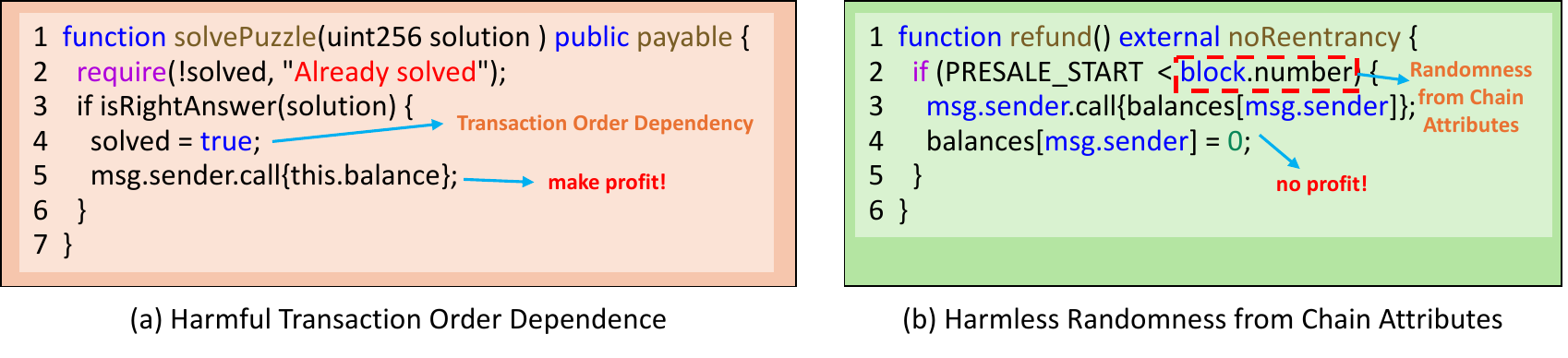}
    \caption{Case study. Two real on-chain contract code snippets, each flagged as vulnerable by detection tools.}
    \label{fig:case_study}
\end{figure}
\vspace{-0.3cm}

\subsection{Case Study}
\label{subsec_case_study}

% Detection tools typically focus on vulnerability characteristics to ensure completeness, without considering the existence or impact of vulnerabilities. Consequently, users need to spend time and effort to verify the correctness of the results. For example, Fig.~\ref{fig:case_study} shows the code snippets of three contracts. Existing detection methods, e.g., Mythril~\cite{mythril}, Confuzzius~\cite{confuzzius}, and Zeus~\cite{zeus}, identify them as containing vulnerabilities. In detail, Mythril flags \texttt{withdraw} in Fig.~\ref{fig:case_study}~(a) as containing Randomness from Chain Attributes. Meanwhile, Confuzzius considers \texttt{sell} in Fig.~\ref{fig:case_study}~(b) to have Unprotected Ether Withdrawal. Besides, Zeus marks \texttt{solvePuzzle} in Fig.~\ref{fig:case_study}~(c) as having Transaction Order Dependence. However, only \texttt{solvePuzzle} can be attacked in the real world.
In Fig.~\ref{fig:case_study}(a), function \texttt{solvePuzzle} contains a Transaction Order Dependence. Specifically, when a user discovers the correct \texttt{solution} and attempts to claim the reward via a transaction, an attacker can front-run the user by sending the transaction with identical parameters. During PoC generation, the LLM tends to place the attacker’s actions after the user’s, causing the PoC to fail \system{}'s profitability check. In this case, \system{} applies the primitive operation \texttt{change\_order} to guide the LLM in swapping the order of the attacker and user actions. This successfully triggers a profitable exploit and \system{} confirms the vulnerability’s severity. In Fig.~\ref{fig:case_study}(b), line 2 relies on a block attribute and is thus categorized as having a Randomness from Chain Attributes. However, \system{} observes that the generated PoC consistently fails to produce profit, and concludes that this is a false alarm, effectively filtering it out.

\section{Discussion}
\label{sec:discussion}

\subsection{Discussion}
\label{subsec:discussion}
\para{Philosophy of Manually Check} 
\system{} uses Manually Check to enhance the effectiveness. Specifically, rather than forcing a definitive judgment on every vulnerability, Manually Check can enhance the interpretability and credibility of \system{}. Besides, Manually Check category enables \system{} to clearly define the boundary between vulnerabilities and false alarms, allowing for fine-grained classification and better vulnerability management for developers and auditors.
% Manually Check enables a more conservative analysis when sufficient evidence is lacking.

% \para{Verifying in off-chain environment}
% Although off-chain testing introduces several challenges, we argue that conducting contract deployment and testing within PoCs is more valuable. This approach offers greater applicability. For instance, \system{} can assess vulnerabilities in contracts that are still under development. Moreover, when using \system{}, developers and auditors can supply additional information to facilitate more targeted testing. This includes providing specific address parameters to support contract deployment or accurate token prices to assist \system{} in evaluating the profitability of an exploit.

\para{Suboptimal attack practices}
The PoCs generated by \system{} may not represent the optimal attack paths. For example, they might include redundant user actions or fail to yield the maximum possible profit. However, the primary goal of \system{} is to determine whether a vulnerability is triggerable and profitable, rather than to identify the most efficient exploitation strategy. Moreover, since contract states can evolve over time, it is inherently difficult to assert how much profit can be obtained.

\para{Testing with cheatcodes}
\system{} uses cheatcodes for pre-attack setup. For example, using \texttt{deal} to provide sufficient funds to the attacker and the contract, rather than acquiring capital through real-world mechanisms like flash loans. This approach simplifies the PoC workflow and increases the likelihood of successful execution. Moreover, using cheatcodes to simulate realistic attack conditions is a common and accepted practice in smart contract vulnerability PoC generation~\cite{defihacklab}.

\subsection{Threats to Validity}
\label{subsec:threats}
\para{Internal validity}
Since \system{} relies on contract source code for analysis and PoC generation, it is unable to handle contracts without available source code. However, this limitation is mitigated by two factors. First, most users tend to adopt open-source contracts and DApps, and a large portion of them are publicly available~\cite{nyx}. Second, \system{} is designed for developers and auditors, who typically have access to the source code. Additionally, \system{} currently lacks robust support for contracts that require complex constructor parameters or involve multi-token interactions, which may lead to inaccurate vulnerability assessments. As future work, we will explore how to leverage already-deployed contracts within a DApp to infer complex parameters and token exchange rate. Moreover, the limited mathematical capabilities of LLMs hinder \system{}’s ability to handle complex arithmetic, leading to false negatives. In future work, drawing inspiration from agent-based approaches, we plan to explore how to integrate external tools to enhance \system{}’s ability. Furthermore, although we enhanced the performance of \system{} through program analysis and carefully crafted prompts, different models still yield varying performance across vulnerability types. In practice, aggregating results from multiple models can further enhance overall effectiveness.
% In practical use, users can mitigate these limitations by supplying \system{} with additional information tailored to their needs. 

\para{External validity}
Currently, \system{} supports vulnerability validation only for Solidity~\cite{solidity} and Vyper~\cite{vyper} smart contracts. Specifically, this is because Foundry, the underlying execution framework, only supports these two languages. While other smart contract languages exist, such as Move~\cite{move}, we believe the impact of this limitation is minimal. In detail, Solidity and Vyper are the first and fourth most widely used smart contract languages, respectively, and together manage over 88\% of the total assets~\cite{top_language}. Besides, SmartBugs contains only a limited number of vulnerabilities, and its data has been used in LLM training, which could bias the results. However, on the one hand, sample size does not affect the effectiveness of \system{}, as profit-based severity assessment is general and \system{} derives representative characteristics from existing detectors and attack patterns. On the other hand, consistent with prior study~\cite{A1}, our experiments indicate that even when the information is provided, current models still struggle to generate valid PoCs. In future work, we will explore vulnerability injection~\cite{ghaleb2020effective} to enable a more comprehensive and fair evaluation. Additionally, due to the performance bottleneck caused by manually constructing the triggering rules, \system{} currently covers only the five most common and financially exploitable vulnerabilities. Extending it to additional vulnerabilities requires extra engineering effort. In future work, we will explore automating the construction process. However, this limitation does not prevent \system{} from adapting to new vulnerabilities. As a vulnerability validation tool, \system{} can feasibly and inexpensively derive new patterns from existing exploits and detection tools.
% \system{} currently supports the evaluation of only five types of vulnerabilities. These are selected because they represent the most common and financially exploitable vulnerabilities, and are widely covered by existing detection tools. In future work, we plan to extend \system{} to support a broader range of vulnerability types.

\section{Related Work}
\label{sec:related_work}
\para{Smart contract vulnerability detection}
A variety of methods have been proposed for smart contract vulnerability detection. For example, fuzzing generates inputs based on the contract’s ABI and observes execution behavior and state changes to identify vulnerabilities. Tools such as Harvey~\cite{harvey} and sfuzz~\cite{sFuzz} apply greybox fuzzing for vulnerability detection, while IcyChecker~\cite{icychecker}, InsFinder~\cite{insfinder}, and SmartReco~\cite{smartreco} leverage rich on-chain states to enable more efficient discovery of vulnerabilities. In contrast, static analysis detects vulnerabilities by examining contract source code or bytecode. For instance, DeFiTainter~\cite{defiTainter} performs static taint analysis at the bytecode level to detect price manipulation vulnerabilities, and Sailfish~\cite{sailfish} uses static analysis on source code to identify state inconsistency issues. With the rise of LLMs, many studies have integrated LLMs into vulnerability detection. For example, GPTScan~\cite{gptscan} combines LLMs with static analysis to detect logical vulnerabilities in smart contracts, while PropertyGPT~\cite{propertygpt} applies LLMs to assist in formal verification of smart contracts. However, most of these detection approaches focus solely on identifying the presence of vulnerabilities, while neglecting their exploitability. As a result, users still need to spend significant effort analyzing the impact of the reported vulnerabilities.

\para{Smart contract automatic exploit generation}
Automatic exploit generation aims to demonstrate the existence of vulnerabilities by triggering them. One common approach is to use symbolic execution to explore potential vulnerability paths. For example, teEther~\cite{teether} performs symbolic execution at the bytecode level to discover exploit paths, while FlashSyn~\cite{flashsyn} combines symbolic execution with counterexample synthesis to construct flash loan attack paths. Besides, ExGen~\cite{exgen} generates symbolic attack contracts and verifies their feasibility via symbolic execution. However, symbolic execution suffers from the path-explosion problem, which limits its applicability. Another widely used technique is fuzzing. ETHPLOIT~\cite{ethploit} applies fuzzing to search for executable paths that can trigger vulnerabilities. AdvScanner~\cite{advscanner} leverages LLM to generate and execute attack contracts targeting Reentrancy. However, these existing efforts primarily focus on detecting whether a vulnerability can be triggered, leading to lots of false alarms. In addition, as LLMs become increasingly capable of code generation and reasoning, some work leverages LLMs to generate PoCs and explore vulnerabilities from multiple perspectives. For example, XPLOGEN~\cite{xplogen} feeds the LLM with known vulnerable contracts and exploits, and further incorporates DCR graphs to generate new vulnerable contracts and corresponding exploits. Besides, A1~\cite{A1} generates verifiable PoCs for vulnerability detection based on on-chain states and inline commands. In contrast, \system{} uses a customized PoC execution environment to perform off-chain PoC generation and validation from scratch. Moreover, with the support of triggerability and profitability analysis, as well as feedback-driven refinement, \system{} can effectively distinguish whether a reported vulnerability is truly exploitable, which is substantially more challenging.
% In contrast, \system{} focuses on validating the exploitability of vulnerabilities—a significantly more challenging task. This involves not only confirming the triggerability of a vulnerability but also demonstrating its impact. 

\section{Conclusion}
\label{sec:conclusion}
In this paper, we present \system{}, a specialized tool for validating exploitable smart contract vulnerabilities. Specifically, \system{} synthesizes PoCs guided by vulnerability paths and verifies their validity based on triggerability and profitability analysis. Furthermore, \system{} introduces a feedback-driven exploration method to increase the possibility of validating vulnerability. Evaluation demonstrates that \system{} is effective in identifying exploitable vulnerabilities and eliminating false alarms. Moreover, it can help state-of-the-art tools in reducing false positives. Besides, \system{} demonstrates strong performance in terms of both time efficiency and cost effectiveness.

\section{Data Availability}
\label{sec:data_availability}
To promote smart contract security development, we release the \system{}. These resources can be accessed at https://github.com/jwzhang-zzz/V2E.

%% file: tables/vulnerability_trigger.tex
\begin{table}[htbp]
\centering
\caption{Vulnerability trigger rules used in \system{}. }
\vspace{-0.2cm}
\label{tab:detection_rules}
% Note that all rules ignore operations and state changes resulting from cheatcodes and test environment initialization.
\resizebox{0.8\textwidth}{!}{

\begin{tabular}{l|l}
\hline
\textbf{Vulnerability}                    & \textbf{Trigger Rules} \\ \hline
Unprotected Ether Withdrawal     & $Transfer(tx.origin)$               \\ \hline
Unprotected Selfdestruct         & $op == \texttt{Selfdestruct}$            \\ \hline
Reentrancy                       & $\exists c_j \in c_i, i \neq j \&\& func(c_i) == func(c_j)$               \\ \hline
Transaction Order Dependence     & $write(slot) \in c_i, read(slot) \in c_j \&\&j>i$               \\ \hline

Randomness from Chain Attributes & $ op \in \{\texttt{Timestamp},\texttt{Blockhash},\texttt{Number}\}$             \\ \hline

\end{tabular}

}

\end{table}

%% file: algorithm/execution_fail_analysis.tex
\begin{center}
\begin{minipage}{0.8\textwidth}  % 设置固定宽度的minipage并居中
    \begin{algorithm}[H]
        % \footnotesize
        \caption{Failed PoC Update Algorithm.}
        \label{algor:execution_fail}
        \renewcommand{\algorithmicrequire}{\textbf{Input:}}
        \renewcommand{\algorithmicensure}{\textbf{Output:}}
        
        \begin{algorithmic}[1]
            \REQUIRE Executed bytecode list $L_{b}$, failed PoC $PoC_{fail}$, Contract source code $s$  %%input
            \ENSURE Updated PoC $PoC_{update}$    %%output
            
            \STATE  $msg_{revert}$ := analyzeOutput($L_{b}$)
            \FOR{$b$ in $L_{b}$.reverse()}
                \STATE $sourceMapping$ := compile($s$)
                \STATE $sourceCode$ := $sourceMapping$.convert($b$)
                \IF{type($sourceCode$) in \{Check, State Update\}}
                    \STATE  $loc$ := $source\_code$ // code where execution failed
                    \STATE  $func$ := searchFunction($loc$) // function where execution failed
                \ENDIF
            \ENDFOR
            \STATE  $PoC_{update}$ := failedPocUpdate($PoC_{fail}$, $loc$, $func$, $msg_{revert}$)
        \end{algorithmic}
    \end{algorithm}
\end{minipage}
\end{center}

%% file: tables/primitive_operation.tex
\begin{table}[htbp]
\centering
\caption{Primitive operation used in \system{}. TA=Trigger analysis. PA=Profit analysis.}
\label{tab:primitive_operation}
\resizebox{0.98\textwidth}{15mm}{
% \usepackage{multirow}
\begin{tabular}{l|l|lllll|ll}
\hline
\multicolumn{1}{c|}{}                                & \multicolumn{1}{c|}{}                                               & \multicolumn{5}{c|}{\textbf{Vulnerability}}                                                                                                                                                                                      & \multicolumn{2}{c}{\textbf{Stage}}                                              \\ \cline{3-9} 
\multicolumn{1}{l|}{\multirow{-2}{*}{\textbf{Name}}} & \multicolumn{1}{c|}{\multirow{-2}{*}{\textbf{Primitive Operation}}} & \multicolumn{1}{c|}{\textbf{RE}}              & \multicolumn{1}{c|}{\textbf{RCA}}             & \multicolumn{1}{c|}{\textbf{TOD}}             & \multicolumn{1}{c|}{\textbf{UEW}}             & \multicolumn{1}{c|}{\textbf{US}} & \multicolumn{1}{c|}{\textbf{TA}}              & \multicolumn{1}{c}{\textbf{PA}} \\ \hline
add\_user                                            & Add additional users and let them invoke \{function\}.         & \multicolumn{1}{l|}{}                         & \multicolumn{1}{l|}{\cellcolor[HTML]{9B9B9B}} & \multicolumn{1}{l|}{\cellcolor[HTML]{9B9B9B}} & \multicolumn{1}{l|}{\cellcolor[HTML]{9B9B9B}} &                                  & \multicolumn{1}{l|}{}                         & \cellcolor[HTML]{9B9B9B}        \\ \hline
change\_invoker                                     & Change the invoker of \{contract\}.                        & \multicolumn{1}{l|}{\cellcolor[HTML]{9B9B9B}} & \multicolumn{1}{l|}{}                         & \multicolumn{1}{l|}{}                         & \multicolumn{1}{l|}{\cellcolor[HTML]{9B9B9B}} & \cellcolor[HTML]{9B9B9B}         & \multicolumn{1}{l|}{\cellcolor[HTML]{9B9B9B}} &                                 \\ \hline
change\_order                                        & Swap the order of attackers and other users.                        & \multicolumn{1}{l|}{}                         & \multicolumn{1}{l|}{}                         & \multicolumn{1}{l|}{\cellcolor[HTML]{9B9B9B}} & \multicolumn{1}{l|}{\cellcolor[HTML]{9B9B9B}} &                                  & \multicolumn{1}{l|}{}                         & \cellcolor[HTML]{9B9B9B}        \\ \hline
modify\_block                                        & Modify the \{block\_attribute\}.                                    & \multicolumn{1}{l|}{}                         & \multicolumn{1}{l|}{\cellcolor[HTML]{9B9B9B}} & \multicolumn{1}{l|}{}                         & \multicolumn{1}{l|}{}                         &                                  & \multicolumn{1}{l|}{\cellcolor[HTML]{9B9B9B}} &                                 \\ \hline
change\_argument                                     & Change the argument or value of \{function\}.                       & \multicolumn{1}{l|}{\cellcolor[HTML]{9B9B9B}} & \multicolumn{1}{l|}{\cellcolor[HTML]{9B9B9B}} & \multicolumn{1}{l|}{\cellcolor[HTML]{9B9B9B}} & \multicolumn{1}{l|}{\cellcolor[HTML]{9B9B9B}} & \cellcolor[HTML]{9B9B9B}         & \multicolumn{1}{l|}{\cellcolor[HTML]{9B9B9B}} & \cellcolor[HTML]{9B9B9B}        \\ \hline
\end{tabular}
}

\end{table}

%% file: tables/dataset.tex
\begin{table}[htbp]
\centering
\caption{The distribution of reported vulnerabilities in our dataset. We manually verify and classify each vulnerability based on whether it is indeed exploitable.}
\label{tab:dataset}
\resizebox{0.7\textwidth}{!}{
\begin{tabular}{ll|ccccc|c}
\hline
\multicolumn{2}{c|}{}                                                                               & \textbf{UEW} & \textbf{US} & \textbf{RE} & \textbf{TOD} & \textbf{RCA} & \textbf{Sum} \\ \hline
\multicolumn{1}{c|}{\multirow{2}{*}{\textbf{$\boldsymbol{D}_{manual}$}}} & \textbf{Exploitable}     & 8            & 1           & 29          & 4            & 8            & 50           \\ \cline{2-8} 
\multicolumn{1}{c|}{}                                                    & \textbf{Non-Exploitable} & 7            & 0           & 2           & 1            & 4            & 14           \\ \hline
\multicolumn{1}{c|}{\multirow{2}{*}{$\boldsymbol{D}_{onchain}$}}         & \textbf{Exploitable}     & 0            & 1           & 6           & 45           & 22           & 74           \\ \cline{2-8} 
\multicolumn{1}{c|}{}                                                    & \textbf{Non-Exploitable} & 1            & 1           & 55          & 31           & 38           & 126          \\ \hline
\multicolumn{2}{l|}{\textbf{Total}}                                                                 & 16           & 3           & 92          & 81           & 72           & 264          \\ \hline
\end{tabular}
}

\end{table}

%% file: tables/profitable_analysis.tex
\begin{table}[htbp]
\centering
\caption{The verification results of \system{}. MC (Exploit | Non-Exploit) refers to instances that require further manual check (inspection).}
\label{tab:verifying_vulnerability}
\resizebox{0.6\textwidth}{!}{

\begin{tabular}{l|ccccc|ccccc}
\hline
\multirow{2}{*}{} & \multicolumn{5}{c|}{$\boldsymbol{D}_{manual}$}                               & \multicolumn{5}{c}{\textbf{$\boldsymbol{D}_{onchain}$}}                              \\ \cline{2-11} 
                  & \textbf{TP} & \textbf{TN} & \textbf{FP} & \textbf{FN} & \textbf{MC} & \textbf{TP} & \textbf{TN} & \textbf{FP} & \textbf{FN} & \textbf{MC} \\ \hline
\textbf{UEW}      & 6           & 5           & 1           & 0           & 2 | 1       & 0           & 1           & 0           & 0           & 0 | 0       \\ \hline
\textbf{US}       & 1           & 0           & 0           & 0           & 0 | 0       & 1           & 1           & 0           & 0           & 0 | 0       \\ \hline
\textbf{RE}       & 28          & 1           & 0           & 1           & 0 | 1       & 6           & 28          & 2           & 0           & 0 | 25      \\ \hline
\textbf{TOD}      & 3           & 1           & 0           & 0           & 1 | 0       & 41          & 8           & 4           & 1           & 3 | 19      \\ \hline
\textbf{RCA}      & 6           & 3           & 0           & 0           & 2 | 1       & 10          & 23          & 2           & 5           & 7 | 13      \\ \hline
\textbf{Total}    & 44          & 10          & 1           & 1           & 5 | 3       & 58          & 61          & 8           & 6           & 10 | 57     \\ \hline
\end{tabular}
}

\end{table}

%% file: tables/baseline.tex
\begin{table}[htbp]
\centering

\caption{The validation consequences compared \system{} with $A1_{re}$ and $LLM_{multi}$.}
\label{tab:baseline}
%\resizebox{0.99\textwidth}{1.25cm}
{
\begin{tabular}{l|ccc|ccc|cccccc}
\hline
\multicolumn{1}{c|}{\multirow{2}{*}{}} & \multicolumn{3}{c|}{\textbf{$\boldsymbol{D}_{manual}$}} & \multicolumn{3}{c|}{\textbf{$\boldsymbol{D}_{onchain}$}} & \multicolumn{6}{c}{\textbf{Total}}                                                    \\ \cline{2-13} 
\multicolumn{1}{c|}{}                  & \textbf{TP}       & \textbf{FP}      & \textbf{FN}      & \textbf{TP}       & \textbf{FP}       & \textbf{FN}      & \textbf{TP} & \textbf{FP} & \textbf{FN} & \textbf{Pre.} & \textbf{Rec.} & \textbf{F1} \\ \hline
$\boldsymbol{A1}_{re}$                 & 34                & 8                & 61               & 29                & 51                & 45               & 63          & 59          & 61          & 51.6\%        & 50.8\%        & 51.2\%      \\ \hline
\textbf{$\boldsymbol{LLM}_{multi}$}    & 31                & 6                & 14               & 12                & 46                & 60               & 43          & 52          & 74          & 45.2\%        & 34.7\%        & 39.3\%      \\ \hline
\textbf{\system{}}                           & 44                & 1                & 1                & 58                & 8                 & 6                & 102         & 9           & 7           & 91.9\%        & 82.3\%        & 86.8\%      \\ \hline
\end{tabular}
}

\end{table}

%% file: tables/false_alarms.tex
\begin{table}[htbp]
\centering
\caption{The results of using \system{} to eliminate false alarms in Slither, Mythril, and Confuzzius.}
\vspace{-0.2cm}
\label{tab:false_alarm}
\resizebox{0.8\textwidth}{!}{
% % \begin{tabular}{cc|ccccc|cc|ccccc}
% % \hline
% % \multicolumn{2}{c|}{\textbf{Slither}} & \multicolumn{5}{c|}{\textbf{Slither+Agen}}                          & \multicolumn{2}{c|}{\textbf{Mythril}} & \multicolumn{5}{c}{\textbf{Mythril+Agen}}                           \\ \hline
% % \textbf{TP}       & \textbf{FP}       & \textbf{TP} & \textbf{TN} & \textbf{FP} & \textbf{FN} & \textbf{NS} & \textbf{TP}       & \textbf{FP}       & \textbf{TP} & \textbf{TN} & \textbf{FP} & \textbf{FN} & \textbf{NS} \\ \hline
% % 37                & 192               & 27          & 102         & 2           & 4           & 6 (88)      & 39                & 78                & 28          & 50          & 3           & 2           & 9 (25)     
% % \end{tabular}
\begin{tabular}{l|ccc|ccccc}
\hline
\multirow{2}{*}{}   & \multicolumn{3}{c|}{\textbf{Original Tool (Baseline)}} & \multicolumn{5}{c}{\textbf{Verification with \system{}}}                                \\ \cline{2-9} 
                    & \textbf{TP}     & \textbf{FP}    & \textbf{FP Rate}    & \textbf{TP} & \textbf{FP} & \textbf{FN} & \textbf{FP Rate} & \textbf{FP Elimination} \\ \hline
\textbf{Slither}    & 37              & 192            & 83.8\%              & 27          & 2           & 4           & 6.9\%            & -76.9\%              \\ \hline
\textbf{Mythril}    & 39              & 78             & 66.6\%              & 28          & 3           & 2           & 9.7\%            & -56.9\%              \\ \hline
\textbf{Confuzzius} & 46              & 106            & 69.7\%              & 41          & 2           & 2           & 4.7\%            & -65\%                \\ \hline
\end{tabular}
}

\end{table}

%% file: tables/different_llm.tex
\begin{table}[htbp]
\centering
\caption{The vulnerability severity verification results under different LLMs.}
\vspace{-0.2cm}
\label{tab:different_llms}
\resizebox{0.65\textwidth}{!}{
\begin{tabular}{l|cccccc}
\hline
                         & \textbf{TP} & \textbf{FP} & \textbf{FN} & \multicolumn{1}{l}{{ \textbf{Pre.}}} & \multicolumn{1}{l}{{ \textbf{Rec.}}} & { \textbf{F1}} \\ \hline
\textbf{GPT-4o}          & 102         & 9           & 7           & { 91.9\%}                                & { 82.3\%}                              & { 86.8\%}      \\ \hline
\textbf{DeepSeek R1}     & 74          & 7           & 31          & { 91.4\%}                                & { 59.7\%}                              & { 72.2\%}      \\ \hline
\textbf{Claude Sonnet 4} & 95          & 11          & 19          & { 89.6\%}                                & { 76.6\%}                              & { 82.6\%}      \\ \hline
\end{tabular}
}

\end{table}

%% file: tables/ablation_study.tex
\begin{table}[htbp]
\centering
\caption{Detection results of \system{}, without vulnerability path analysis, without exploit and profit-driven analysis, and without feedback-driven PoC fine-tuning based on $D_{manual}$ and $D_{onchain}$.}
\vspace{-0.2cm}
\label{tab:ablation_study}
\resizebox{0.7\textwidth}{!}{
\begin{tabular}{l|ccccc}
\hline
                                                                    & \textbf{TP}               & \textbf{TN}               & \textbf{FP}               & \textbf{FN}               & \textbf{MC}               \\ \hline
\textbf{Without vulnerability path analysis}                        & 29                        & 46                        & 10                        & 45                        & 134                       \\ \hline
{\textbf{Without trigger rule-based analysis}} & { 44} & { 59} & { 91} & { 65} & { 5}  \\ \hline
{\textbf{Without LLM-based profit analysis}}   & { 62} & { 63} & { 32} & { 79} & { 28} \\ \hline
\textbf{Without feedback-driven PoC refinement}                     & 37                        & 58                        & 9                         & 45                        & 115                       \\ \hline
\textbf{\system{}}                                     & 102                       & 71                        & 9                         & 7                         & 75                        \\ \hline
\end{tabular}
}

\end{table}